\documentclass[useAMS,usenatbib]{mn2e}
\usepackage{graphicx}
\usepackage{epsfig}
\usepackage{rotating}
\usepackage{amssymb}

\title[Tracing substructure in A1300]{In the whirlpool's coils: tracing substructure from combined optical/X-ray data in the galaxy cluster A1300}
\author[F. Ziparo et al.]{F. Ziparo,$^{1}$\thanks{E-mail:
felicia.ziparo@mpe.mpg.de} F.G. Braglia,$^{2}$ D. Pierini,\thanks{Visiting astronomer at MPE} A. Finoguenov,$^{1}$ H. B\"{o}hringer$^{1}$ and   \newauthor A. Bongiorno$^{1}$\\
$^{1}$Max-Planck-Institut f\"{u}r extraterrestrische Physik, Giessenbachstra\ss e 1, 85748 Garching bei M\"{u}nchen, Germany\\
$^{2}$Imperial College London, Blackett Laboratory, Prince Consort Road, London SW7 2AZ, United Kingdom}
\begin{document}

\date{Accepted in MNRAS 2011 November 16; Received 2011 October 21; in original form 2011 July 29}

\pagerange{\pageref{firstpage}--\pageref{lastpage}} \pubyear{2011}

\maketitle

\label{firstpage}

\begin{abstract}
Structure formation is thought to act via hierarchical mergers and accretion of smaller systems driven by gravity with dark matter dominating the gravitational field. Combining X-ray and optical imaging and spectroscopy provides a powerful approach to the study of the cluster dynamics and mass assembly history.
The REFLEX-DXL sample contains the most X-ray luminous galaxy clusters ($\mathrm{L_X \geq 10^{45}\ erg\ s^{-1}}$) from the REFLEX survey at $\mathrm{z = 0.27-0.31}$.
We present the photometric (WFI) and spectroscopic (VIMOS) data for the DXL cluster RXCJ1131.9\textendash  1955 (Abell 1300); in combination with the existing X-ray data we determine and characterise the substructure of this post-merging system.
We analyse X-ray selected groups in a $\mathrm{30\arcmin \times 30\arcmin}$ region encompassing the cluster in order to study the mass assembly of A1300.
The X-ray surface brightness map of A1300 appears disturbed and exhibits the signature of a forward shock, which is consistent with a previous analysis of radio data. Moreover, we detect a large scale-filament in which the cluster is embedded and several infalling groups. 
Comparison of the whirlpool-like features in the entropy pseudo-map of the intra-cluster medium with the distribution of the cluster members reveals a direct correspondence between the ICM structure and the galaxy distribution. Moreover, comparison with existing simulations allows us to better understand the dynamics of the cluster progenitors and to age date their impact.
A1300 is a complex massive system in which a major merging occurred about 3 Gyr ago and additional minor merging events happen at different times via filaments, that will lead to an increase of the cluster mass of up to 60\% in the next Gyr.
\end{abstract}

\begin{keywords}
Galaxies: clusters: individual: Abell 1300 -- Galaxies: clusters: general --  Galaxies: kinematics and dynamics -- Galaxies: evolution -- Cosmology: observations.
\end{keywords}

\section{Introduction}
Structure formation and evolution are thought to proceed via hierarchical merger events and accretion of smaller units to form larger systems driven by gravity with dark matter (DM) dominating the gravitational field.
Thus, merging plays a key role in driving the build-up of structures on both small and large scales.
Clusters are located at the crossing point of filamentary structures which drive the accretion towards their DM potential wells. Accretion of smaller systems happens preferentially in the filaments and the products are, in turn, accreted by larger clusters (e.g. \citealt{2005Natur.435..629S}).
Investigating massive merging clusters implies having a snapshot of the regions of the Universe where the highest-mass structures are created, which offers constraints and insights to both cosmology and astrophysics. Indeed, the study of galaxy clusters allows not only to highlight the structure of the Universe and to quantify several cosmological parameters but also to study the dynamics and evolution of the baryonic fraction in a deep potential well \citep{2009arXiv0906.4370B}.

In most cases the dynamical state of clusters has been derived from X-ray observations (e.g. \citealt{1992csg..conf...49J} and \citealt{2002ASSL..272...79B}). X-ray studies of merging clusters are useful to provide information on the evolution of the intra-cluster medium (ICM, e.g. \citealt{2002ASSL..272....1S}), for example on how the ICM is heated to the high observed temperatures and how entropy structure is generated (\citealt{2001ApJ...546...63T}, \citealt{2010A&A...511A..85P}), in particular through the effect of shocks (e.g. \citealt{2007PhR...443....1M} for a review).

Merging events are detected by substructures in the surface brightness distribution of X-ray images (e.g. \citealt{2001A&A...378..408S}). By their characterisation we are able to study the cluster dynamics and mass assembly history. For example, combining X-ray and spectroscopic information \cite{2011ApJ...728...27O} analysed the merging cluster Abell~2744. The identification of substructures was crucial to determine the merging direction and approximately establish the status of the core passage.
However, accretion events can be also identified by considering all infalling groups in the proximity of a cluster \citep{2009MNRAS.400..937M} and deriving their properties. 
These groups can be detected through the X-ray emission of their gas (e.g. \citealt{2000ARA&A..38..289M}) and through the distribution of galaxies and optical light.
Direct detection of X-ray emission from groups is not always feasible: indeed, small associations of a few galaxies are normally undetectable in X-rays \citep{2009ApJ...704..564F} due to their weak emission, and even richer groups, once captured by a massive cluster, are likely to lose almost all their gas. Nevertheless, it is likely that the galaxies of accreted groups will give rise to significant galaxy over-densities within the parent cluster \citep{2006PASP..118..517B}. Combining these two complementary approaches makes it possible to explore the history of formation and evolution of clusters, where several snapshots of the accretion pattern are highlighted just by observing one cluster out to several times its radius.

In order to understand the role of merging in the evolution of clusters with similar mass and the connection with their large scale structure (LSS) environment we need a sample of galaxy clusters with small mass range and different dynamical states.
With this purpose we have selected a statistically complete cluster sample drawn from the ROSAT ESO Flux Limited X-ray (REFLEX) survey \citep{2001A&A...369..826B}.
The 13 distant X-ray luminous (DXL, see e.g. \citealt{2004astro.ph..2533Z} and \citealt{2005AdSpR..36..667Z} for details) clusters have luminosity $\mathrm{L_X^{bol}=0.5 - 4\times 10^{45} \ erg \ s^{-1}}$, masses $\mathrm{M_{500}= 0.5- 1.1\times 10^{15}M_\odot}$\footnote{$\mathrm{M_\Delta}$ (where $\mathrm{\Delta=500,200}$) is defined as $\mathrm{M_\Delta=(4 \pi/3) \Delta \rho_c R_{\Delta}^3}$ where $\mathrm{R_\Delta}$  is the radius at which the density of a cluster is equal to $\Delta$ times the critical density of the Universe ($\mathrm{\rho_c}$). Throughout our analysis we adopt the X-ray estimate of $\mathrm{R_\Delta}$ (unless it is otherwise specified).} and are located within a narrow redshift interval ($\mathrm{z=0.27-0.31}$). 
The DXL sample is a powerful instrument to investigate the mass assembly of the clusters and the evolution of galaxies therein, together with the exploration of the link between large-scale structure, substructure and galaxy population.
Moreover, this snapshot of the Universe is comparable to N-body simulations including hydrodynamics (e.g. \citealt{2009arXiv0906.4370B}, \citealt{2011ApJ...728...54Z}), allowing us to better investigate the physics that regulates cluster evolution across the cosmic web. For instance, one can understand the variation of the sub-halo mass function, traced by galaxies and the amount of substructure in clusters; estimate time scales from the dynamical state of the gas; and understand the physical processes that drive its behaviour. In fact, DXL can be represented as a sequence of cluster dynamical states, starting from early stages of merging events (including several components of different mass) towards strong cool core clusters.

Most of the detailed X-ray analysis of DXL clusters has been already performed ( \citealt{2004astro.ph..2533Z}, \citealt{2004A&A...413...49Z}, \citealt{2005AdSpR..36..667Z}, \citealt{2006A&A...456...55Z}, hereafter referred to as Z06, \citealt{2005A&A...442..827F}). These studies, focused on the ICM, have provided valuable information on the dynamical state, AGN feedback and chemical enrichment of these clusters.

Optical analysis of the DXL sample has started in parallel. In particular, the attention was focused on two clusters with different dynamical states: an ongoing major merger \citep{2007A&A...470..425B} and a relaxed cluster (\citealt{2009A&A...500..947B},  hereafter referred to as B09), in order to compare their kinematics and galaxy distributions with their X-ray properties.
\citet{2007A&A...470..425B} found, associated with the merging cluster A\,2744, two large-scale filaments along which blue galaxies exhibited enhanced star formation activity. 
This study was followed up in B09 who explored the existing link between the fraction of passively evolving galaxies and the assembly state of the cluster.
\citet{2008A&A...483..727P} then suggested, from observations of three DXL clusters, that the intra cluster light has multiple origins, possibly linked to the dynamical state of the cluster.

In this paper we present the results from the study of RXCJ1131.9\textendash  1955 (alias Abell 1300), a post-merging cluster at $\mathrm{z \sim 0.3075}$ with a ``dumbbell'' cD galaxy \citep{1997A&AS..124..283P} at its centre and prominent filaments visible in the galaxy density distribution. The definition of post-merging cluster dates back to \citet{1997A&A...326...34L} who state that this cluster has undergone a merger but the merging phase may be nearly over. We support this statement after the inspection of the shape of the X-ray emission (Fig.~\ref{BCG_fig}), not clearly separated from the dark matter halo, and the displacement observed between the X-ray peak and the Brightest Cluster Galaxy (BCG), as better explained later on.

This paper is organised as follows: in Section 2 we describe our dataset and our preliminary analysis; in Section 3 we give a morphological overview of the cluster and its large scale structure using optical (photometric and spectroscopic) and X-ray analyses; in Section 4 we discuss our results comparing them with simulations and previous works; we draw our conclusions in Section 5. Throughout our analysis we adopt the AB magnitude system (unless otherwise specified) and the following cosmological values: $\mathrm{H_0=70} \mathrm{\ km\ s^{-1}\ Mpc^{-1}}$, $\Omega_\mathrm{M}=0.3$ and $\Omega_\mathrm{\Lambda}=0.7$. At the cluster distance 1$\arcmin$ corresponds to 270 kpc. In all figures hereafter North is up and East to the left, unless otherwise specified.

\section{Observations, Data Reduction and Analysis}
\subsection{Wide-field Imaging}
Optical photometry was carried out using the Wide Field Imager (WFI, \citealt{1999Msngr..95...15B}) mounted on the Cassegrain focus of the ESO/MPG-2.2~m telescope at La Silla, Chile. 
The data presented here were obtained as part of a heterogeneous programme during MPG observing time in visitor mode (P.I.: B\"{o}hringer). The observations of A1300 in the B, V and R passbands were performed in 2001, between January 27th and February 1st, in photometric conditions. 
They were divided into sequences of 8 dithered sub-exposures for a total exposure time of $\mathrm{3150 \ s}$ for the V band and $\mathrm{3600 \ s}$ for the R and B bands.
Filter curves can be found in \citet{2001A&A...379..740A} and on the web-page of the La Silla Science Operation Team\footnote{http://www.eso.org/sci/facilities/lasilla/instruments/wfi/inst/ filters}. Standard stars were observed in all the four nights: three Landolt fields \citep{1992AJ....104..340L} were targeted for a total of 35-50 standard stars per field (SA98, SA101 and SA104).

The WFI data were reduced using the data reduction system developed for the ESO Imaging Survey (EIS, \citealt{1997Msngr..87...23R}) and its associated EIS/MVM image processing library version 1.0.1 ($Alambic$\footnote{http://www.eso.org/sci/activities/projects/eis/survey\_$ $release. html}). 
For more details on the transformation of raw images into reduced ones see \citet{2008A&A...483..727P}.
\begin{table}
\begin{center}
% use packages: array
\begin{tabular}[t]{cccc}
\hline
Passband & ZP & $k$ & CT \\ 
\hline
B & 24.55$\pm$0.01 & 0.20$\pm$0.01 & 0.24 \\ 

V & 24.15$\pm$0.01 & 0.14$\pm$0.01 & -0.12 \\ 

R & 24.43$\pm$0.01 & 0.09$\pm$0.01 & 0.01 \\
\hline

\end{tabular}
\end{center}
\caption[zp work]{Photometric solutions determined in this work. Column 1 gives the passband, column 2 the zero-points in the Vega magnitude system, column 3 the extinction coefficient and column 4 the colour term. Two of the three quantities are reported with respective errors in the Vega magnitude. These best-fit parameters were obtained from a two-parameter fit from about 850 to 2000 measurements across the WFI field for each passband, the colour term being fixed. }
\label{zp_work}
\end{table}
Source detection and photometry were based on SExtractor \citep{1996A&AS..117..393B} both for standard and science images.
Magnitudes were calibrated to the Johnson-Cousins system using \citet{1992AJ....104..340L} standard stars whose magnitudes were obtained using a 10 arcsec-wide circular aperture, which were adequate as judged from determining the average growth curve of all the measured stars.
Photometric standards were observed over a rather broad range of airmasses, but science frames were taken at the best airmass; in this way we were able to obtain photometric solutions (e.g. zero-points) for the calibration of reduced scientific images by merging the measurements of standard stars for each passband.
The number of non-saturated Landolt stars per field did not allow independent solutions to be
determined for each of the eight chips of WFI. Hence calibration had to rely upon solutions based on measurements taken across all
chips. Although the EIS data reduction system includes a photometric pipeline for the automatic determination of photometric
solutions, these were determined interactively using the IRAF\footnote{IRAF is the Image Reduction and Analysis Facility, a general purpose software system for the reduction and analysis of astronomical data. IRAF is written and supported by the IRAF programming group at the National Optical Astronomy Observatories (NOAO) in Tucson, Arizona. NOAO is operated by the Association of Universities for Research in Astronomy (AURA), Inc. under cooperative agreement with the National Science Foundation.} task $fitparams$.
This choice allows the interactive rejection of individual measurements, stars, and chips. Photometric solutions
with minimum scatter were obtained by a two-parameter linear fit with about 850 photometric points for the B-band, about 900 for the V-band and more than 2000 for the R-band, the atmospheric extinction coefficient in each band being set equal to that listed as the median value obtained by the EIS team\footnote{http://www.eso.org/sci/activities/projects/eis/surveys/readme/ 70000027}.
In general, zero-points and colour terms are consistent with those obtained by the EIS team or by the 2p2 Telescope Team\footnote{http://www.eso.org/sci/facilities/lasilla/instruments/wfi/zero- points}, as can be seen by comparing Tables \ref{zp_work}\textendash \ref{zp_2p2}.

\begin{table}
\begin{center}

\begin{tabular}[t]{cccc}
\hline
Passband & ZP & $k$ & CT \\ 
\hline
B & 24.65 & 0.23 & 0.24 \\ 

V & 24.19 & 0.15 & -0.12 \\ 

R & 24.47 & 0.12 & 0.01 \\
\hline

\end{tabular}
\end{center}
\caption{Median values for all the photometric solutions in the Vega magnitude system based on three parameter
fits obtained by the ESO Deep Public Survey (DPS) team.}
\label{zp_eis}
\end{table}

\begin{table}
\begin{center}

\begin{tabular}[t]{cccc}
\hline
Passband & ZP & $k$ & CT \\ 
\hline
B & 24.81$\pm$0.05 & 0.22$\pm$0.015 & 0.25$\pm$0.01 \\ 

V & 24.15$\pm$0.04 & 0.11$\pm$0.01 & -0.13$\pm$0.01 \\ 

R & 24.47$\pm$0.04 & 0.07$\pm$0.01 & 0.00$\pm$0.00 \\
\hline

\end{tabular}
\end{center}
\caption{Definitive photometric solutions obtained by the
2p2 Telescope Team from observations of standard stars in perfectly
photometric nights, where a bunch of standard star fields were moved
around each chip of WFI. All parameters were fitted simultaneously as free parameters, with good
airmass and colour range, and around 300 stars per fit. The Table below
gives the average solutions in the Vega magnitude system over all chips.}
\label{zp_2p2}
\end{table}

As for science images, source extraction and photometry were obtained after matching the BVR images of each target
to the worst seeing (0.92$\arcsec$ for the V band), using the IRAF task $psfmatch$, and taking into account the weight-maps associated with the individual images, produced by $Alambic$.
A common configuration file was used to produce three catalogues per target, after evaluating the seeing and the zero-point for individual images. The R-band image having the deepest exposure was used as the detection image, where sources are defined
by an area with a minimum number of 5 pixels above a threshold of $1\sigma$ of the background counts.
Source photometry in individual passbands was extracted in fixed circular apertures (between $\mathrm{1.2\arcsec}$ and $\mathrm{10\arcsec}$ in diameter) or in flexible elliptical apertures (Kron-like, \citealt{1980ApJS...43..305K}) with a Kron-factor of 2.5 and a minimum radius of 3.5 pixels.
For our analysis we used the total magnitudes (Kron-like). 
Object magnitudes were corrected for Galactic extinction according to the \citet{1998ApJ...500..525S} galactic reddening
maps (from NASA/IPAC Extragalactic Database, NED) and converted to the AB system according to the response function of the optical system (see \citealt{2004A&A...428..339A}). 
The output catalogues were successively culled of fake sources by hand and by using masks before photometric redshifts were determined.
In addition, stars and galaxies could be safely identified on the basis of their surface brightness profile and optical colours down to
$\mathrm{R = 21.5}$. Fainter than this limit, number counts are dominated by galaxies, so that all detected objects with $\mathrm{R > 21.5}$ are assumed to be galaxies. 
This assumption is supported by several tests that we ran, comparing colours and magnitudes derived with different methods. Information from SExtractor flags was also taken into account in these tests.
\begin{figure}
\includegraphics[width=\hsize]{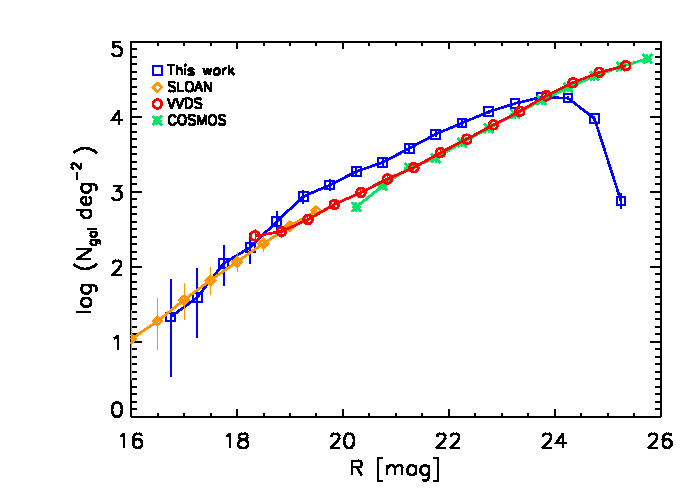}
 \caption{Comparison of number counts for the entire region of A1300 imaged with WFI in this work and three deep fields/wide area surveys: VVDS (red circles), COSMOS (green stars) and SDSS (orange diamonds). All errors, but those of COSMOS (for which errors were available, \citealt{2004AJ....127..180C}), were obtained using the modified Poisson statistics in \citet{1986ApJ...303..336G}. }
 \label{deep_survey}
\end{figure}
Depth and quality of the final catalogues were also determined. 
Galaxy number counts in the observed field were compared with deep number counts from several surveys (i.e. VVDS, VIMOS VLT Deep Survey, \citealt{2003A&A...410...17M}; COSMOS, Cosmic Evolution Survey, \citealt{2004AJ....127..180C}; and SDSS, Sloan Digital Sky Survey, \citealt{2001AJ....122.1104Y}). All errors, but the ones from COSMOS (for which errors were available, \citealt{2004AJ....127..180C}), were obtained using Poissonian statistics from \citet{1986ApJ...303..336G}. 

Our galaxy number counts exceed those obtained from observations of deep fields such as VVDS \citep{2003A&A...410...17M} and COSMOS \citep{2004AJ....127..180C} for $\mathrm{18.5 \leq R \leq 23.5 \ mag}$, as shown in Fig.~\ref{deep_survey}. On the other hand, they begin to drop below the galaxy number counts in deep fields/wide area surveys at $\mathrm{R > 23.5}$, where the number of background galaxies dominates the number of likely cluster members. The bright end is comparable with the number counts of SDSS \citep{2001AJ....122.1104Y}.
Assuming as a completeness limit the magnitude at which the observed counts are equal to 50\% of those in the deep fields/wide area surveys, we thus conclude that our R-band selected catalogues are complete down to $\mathrm{\sim24\ mag_{AB}}$ with respect to the VVDS. 

\begin{figure}
\includegraphics[angle=90,width=\hsize]{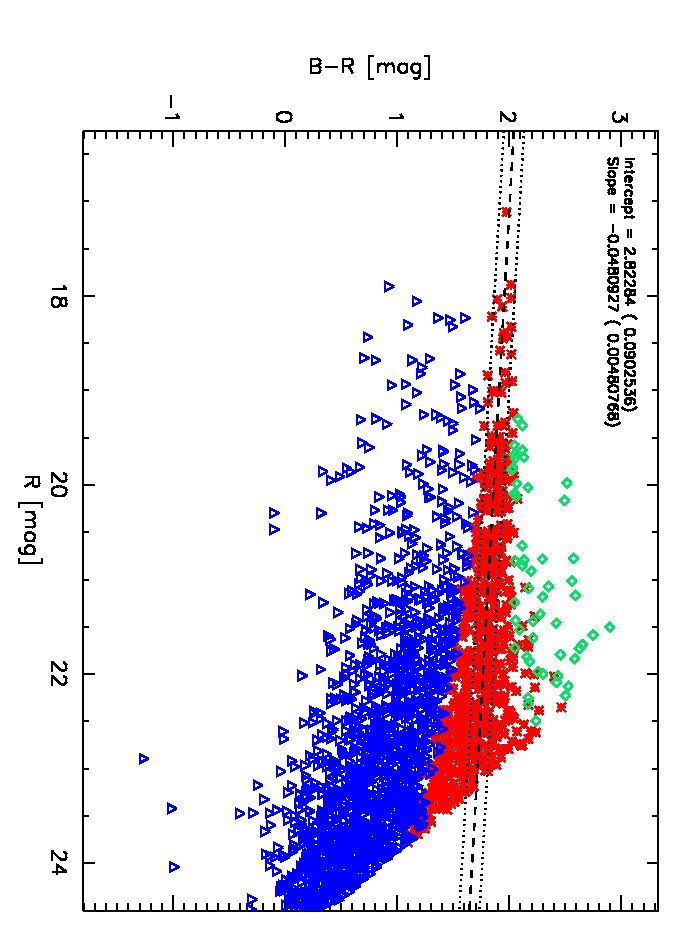}
\caption{Colour-magnitude diagram (B-R versus R) for all galaxies found in the field of A1300 comprised within $\mathrm{R_{200}}$ of the cluster. Red stars represent galaxies defining the red sequence within $\mathrm{3\sigma}$ from the best-fit within the photometric errors, blue triangles show the galaxies bluer then the red sequence of $\mathrm{3\sigma}$ within the photometric errors, green diamonds all galaxies redder than the red sequence galaxies of $\mathrm{3\sigma}$ within the photometric errors. The dashed line represent the red sequence best fit and the dotted line the $\pm 1\sigma$ scatter around the red sequence. The catalogue is cut for the magnitude limit of $\mathrm{B=24.9}$ and $\mathrm{R=24.5}$.}
 \label{CMD}
\end{figure}

We identify and fit the cluster red-sequence from the colour-magnitude diagram (CMD, Fig.~\ref{CMD}) of all the galaxies in the imaged region of A1300 within a cluster-centric distance of $\mathrm{5.66 {\arcmin}}$, corresponding to $\mathrm{R_{200}} = 1.53$ Mpc at the cluster redshift (cf. Z06). The best fit (dashed line in Fig.~\ref{CMD}), obtained through recursive $\mathrm{3\sigma}$ clipping, is described by the linear relation:
\begin{equation}
(B-R)=(2.823 \pm 0.090) -(0.048 \pm 0.005)\times R.
 \label{CMD_equation}
\end{equation} 

The rms scatter around the red sequence fit is 0.09 mag (represented by the dotted lines in Fig.~\ref{CMD}).
We also compiled a CMD for the spectroscopic and photometric members separately as in B09: since these fits did not show a remarkable difference w.r.t. the one in Fig.~\ref{CMD}, we decided to use equation \ref{CMD_equation}, being based on larger statistics.
The identification of the red sequence allowed us to distinguish between red (objects within $\mathrm{3\sigma}$ from the red sequence, likely to be old passively evolving objects) and blue (below the red sequence, likely to be star forming objects) galaxies and to trace their distribution in the cluster field.

\subsection{Spectroscopic Data} 
\label{spectro_data_descr}
Multi-Object Spectroscopy (MOS) was performed between the periods of May 21-23, 2004 and January 12-19, 2005 as part of the ESO GO large programme 169.A-0595 (P.I. B\"ohringer), carried out in visitor and service modes. The main aim of this program was to observe the largest number of galaxies lying in the same area of the sky for seven out of the 13 DXL clusters. 

Low resolution ($\mathrm{R=200}$, LR-Blue grism) spectroscopy was carried out with VIMOS (VIsible Multi-Object Spectrograph, \citealt{2003SPIE.4841.1670L}) mounted on VLT-UT3 at Paranal Observatory (ESO), Chile.
VIMOS is a wide-field imager and multi-object spectrograph operating in the visible (from 3600 to 10000 \AA{}), with an array of 4 identical CCDs with a field of view (FOV) of $\mathrm{7\arcmin \times 8\arcmin}$ each and $0.205\ \arcsec$ pixel scale, separated by a gap between each quadrant of $\mathrm{\sim2\arcmin}$.

\begin{figure}
\includegraphics[width=\hsize]{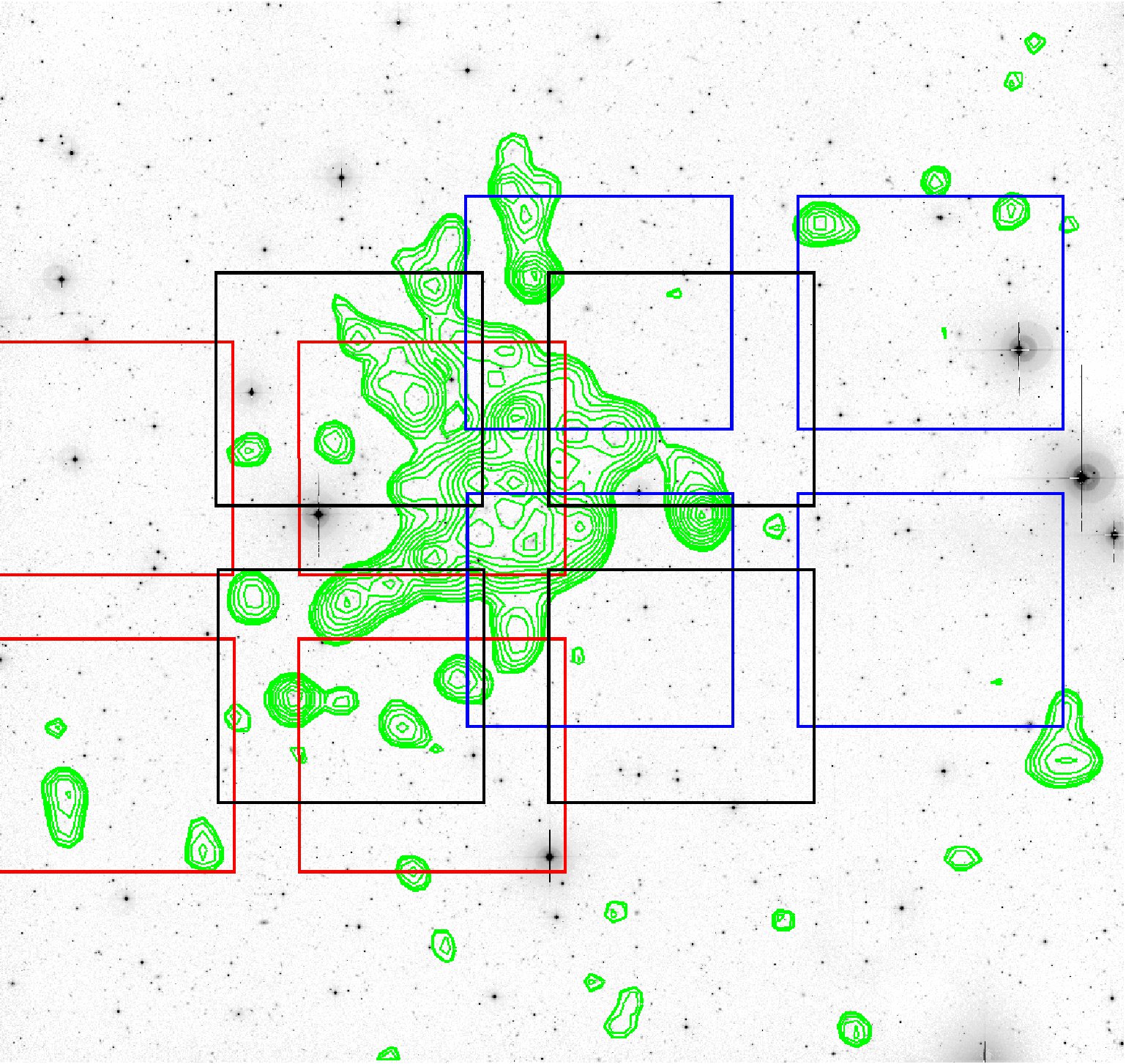}
\caption{R-band image of A1300 on which are overlaid density contours (in green) and the three VIMOS pointings. Each set of four boxes with the same colour represents one VIMOS pointing.}
 \label{vimos_pointings}
\end{figure}

To provide a good coverage of the cluster central region and to extend the analysis to the cluster outskirts we used three pointings that partially overlap in the centre and reach beyond a distance of 4 Mpc in the E-W direction from the cluster X-ray centre (Fig.~\ref{vimos_pointings}).

Objects with $\mathrm{I\leq 22.5}$ were selected from the pre-imaging with VLT-VIMOS, corresponding to a limiting magnitude of approximately $\mathrm{I^\star + 3}$ galaxy at the redshift of the cluster (0.3075, see \citealt{1998ApJ...497..188C}). The pre-imaging was done in the I-band in order to select targets based only on stellar mass (e.g. \citealt{1994ApJS...95..107W}) and avoiding any colour bias. The catalogue on the pre-imaging was produced running SExtractor \citep{1996A&AS..117..393B} which allowed to classify galaxies with $\mathrm{I \leq 20}$ as bright and objects with $\mathrm{20 < I \leq 22.5}$ as faint, in order to observe them with two different masks (prepared using the VMMPS\footnote{http://www.eso.org/sci/observing/phase2/VIMOS/VMMPS. html} tool from ESO) for the same Observation Block (OB).
Slits of $1 \arcsec$ width were used for an expected uncertainty on the observed velocities of $\mathrm{250-300\ km\ s^{-1}}$. Previous works (\citealt{2007A&A...470..425B} and B09) have confirmed these expectations and shown that with these uncertainties it is possible to establish the membership of a galaxy and the global cluster dynamics for the massive clusters of REFLEX\textendash DXL.

At the average redshift of the DXL ($\mathrm{z\sim0.3}$) the LR-Blue grism samples several important spectral features: [OII], [OIII], CaII$_{\mathrm{H+K}}$, $\mathrm{H_\beta}$, $\mathrm{H_\delta}$ lines and the 4000\,\AA \,break. The combination of these features allows us to characterise the spectral type of galaxies (e.g. by the 4000\,\AA \,break), the present star formation rate (e.g. by the [OII] line) and nuclear activity (by the [OII]/[OIII] line ratio). The spectrum in this wavelength range does not suffer from fringing and spectroscopic redshifts up to $\mathrm{z\sim0.8}$ can be determined.

The spectroscopic observations provided about 900 spectra which were reduced using the dedicated software VIPGI\footnote{VIPGI (VIMOS Interactive Pipeline and Graphical Interface, \citealt{2005PASP..117.1284S}). This software was developed by the VIRMOS Consortium to handle the reduction of the VIMOS data for the VVDS (\citealt{2003A&A...410...17M})}. VIPGI allows to calibrate all spectra through a user-friendly interface, applying flat fields corrections, sky subtraction, spectrum extraction and wavelength calibration. Data reduction followed the standard approach, as described in the VIPGI manual and in \citet{2005PASP..117.1284S}: lines were fitted and matched with available line catalogues. Furthermore, we used the template fitting procedure EZ (Easy redshift, \citealt{2010PASP..122..827G}), that allows to fit spectral templates to the continuum when no evident features (e.g. emission lines) were present. Although EZ mainly relies on a $\mathrm{\chi^2}$ template-fitting procedure it allows also to choose the best template after an analysis by eye. We assigned different flags to the redshifts according to their reliability: $flag=0$ was given when it was not possible to assign a redshift, $flag=1$ meant a confidence in the redshift within 25\%, $flag=2$ a reliability of the solution comprised between 25\% and 50\%, $flag=3$ between 50\% and 75\% and $flag=4$ between 75\% and 100\%. In order to ensure reliable cluster memberships we only
use galaxies with spectral flags larger than 2 in the following.
We expanded our sample of spectroscopic redshifts with 96 publicly available ones from NED.
\begin{figure*}
\includegraphics[width=0.48\hsize]{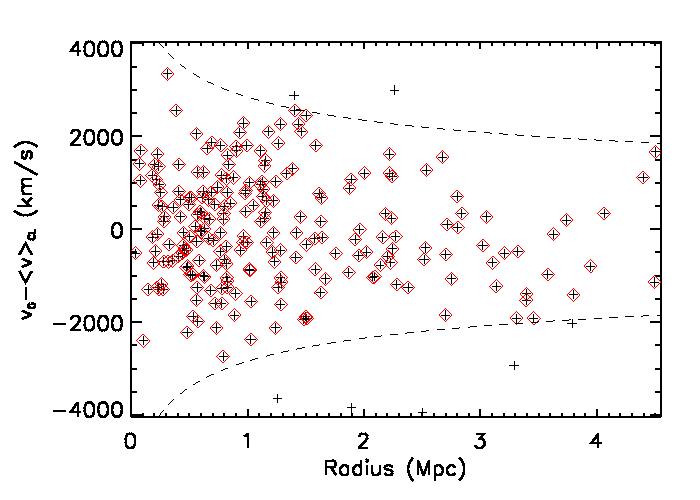}
\includegraphics[width=0.48\hsize]{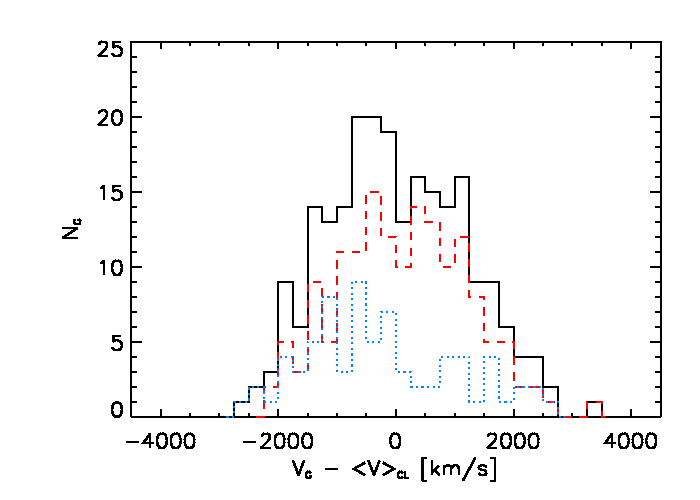}
\caption{Left panel: phase-space diagram of the confirmed 230 spectroscopic cluster members for A1300. Crosses mark the candidate cluster members prior to the interloper rejection, while red diamonds identify the confirmed cluster members. The dashed lines track the caustics defined by \citet{2004ApJ...600..657K} for comparison. Right panel: velocity distribution of the spectroscopic cluster member galaxies. The black solid histogram shows the distribution of all cluster members; the distribution of velocities for blue and red members (as defined from their position in the colour-magnitude diagram, cf. Section \ref{photometric_redshifts_section}) is shown by the dashed red and dotted blue histograms, respectively. }
\label{members}
\end{figure*}

\subsubsection{Cluster membership}
\label{photometric_redshifts_section}
Spectroscopic cluster members were identified using the same combination of techniques summarised by \citet{2006A&A...456...23B}. First, we removed the obvious interlopers by excluding all galaxies with peculiar velocities in excess of $\mathrm{\pm 4000\  km\ s^{-1}}$ from the robust cluster redshift of $0.3048 \pm 0.0044$ (calculated using the \citet{1990AJ....100...32B} biweight estimators for robust mean and scale). Peculiar velocities were corrected for cosmological redshift and velocity errors and set to rest-frame using the standard recipe of \citet{1980A&A....82..322D}. To the remaining galaxies we applied first the weighted gap selection method described by \citet{1993ApJ...404...38G} and then the phase-space rejection criterion of \citet[cf. also \citealt{2004ApJ...600..657K}]{1996MNRAS.279..349D} to identify less evident interlopers. In all our calculations we assumed as cluster centre the peak of the X-ray surface brightness map, which lies within $8\arcsec$ from the BCG.

Our analysis yields a total of 230 dynamically bound galaxies within a projected cluster-centric distance of $\mathrm{4.5\ h_{70}^{-1}\ Mpc}$, with a robust rest-frame velocity dispersion of $\mathrm{987 \pm 101\ km\ s^{-1}}$ (the errors on velocity dispersion and later on mass are derived via jackknife from the catalogue of confirmed cluster members). Fig.~\ref{members} shows the distribution of cluster members in phase-space and their velocity distribution.

The spectroscopic redshifts  enabled us to train the photometric redshift solutions obtained from Le PHARE (PHotometric Analysis for Redshift Estimations, S. Arnouts \& O. Ilbert), a publicly available\footnote{http://www.cfht.hawaii.edu/~arnouts/LEPHARE/cfht\_lephare/ lephare.html} software based on the $\mathrm{\chi^2}$ template-fitting procedure. 
The photometric data in three optical bands (B, V, R) allowed us to trace the Balmer break of galaxies as a function of redshift up to $\mathrm{z \sim 1.5}$. 
We adopt the included Virgo cluster template set \citep{2003A&A...402...37B}, as this yields the highest quality photometric redshifts. This was defined as the smallest achievable fraction of catastrophic failures $\mathrm{\eta=|z_p -z_s|/(1+z_s)>0.15}$ (where $\mathrm{z_p}$ and $\mathrm{z_s}$ are the photometric and spectroscopic redshifts, respectively) combined with the best possible accuracy $\mathrm{\sigma_{\Delta z/(1+z_s)}}$ measured with the normalised median absolute deviation $\mathrm{1.48 \times median(|\Delta z|/(1+z))}$ (as done in \citealt{2006A&A...457..841I}), where $\mathrm{\Delta z = z_p-z_s}$. The value of these parameters for all galaxies (without any selection in magnitude) and for bright galaxies ($\mathrm{R<R^\star +1}$) were respectively: $\mathrm{\eta = 0.21}$, $\mathrm{\sigma_{\Delta z/(1+z_s)} = 0.07}$, $\mathrm{\eta_{Bright} = 0.07}$ and $\mathrm{\sigma_{Bright,\Delta z/(1+z_s)} = 0.03}$. We removed the Blue Compact Dwarf template as this was found to increase the scatter and the number of catastrophic failures.

We use the option AUTO\_ADAPT of Le PHARE to correct our zero-points based on a sub-sample of bright galaxies and then we apply the result to the whole catalogue. We chose the template set and checked whether to apply extinction correction on several templates using modified \citet{2000ApJ...533..682C} attenuation law and several others with the colour excess E(B-V) values ranging between 0 and 0.3 and with a step of 0.05. Eventually, we did not use any extinction correction as we obtained our best result using the provided templates. 
\citet{2009ApJ...690.1236I} implemented an improved method to compute photometric redshifts taking into account the emission line contribution using relations between the UV continuum and the emission line fluxes associated with star-formation
activity (like $\mathrm{[OII],\ [OIII],\ H_\beta, \ H_\alpha}$).
The authors compared the template curves with and without emission lines and found that the expected line fluxes can change up to 0.4 mag in the colour. Therefore, we decided to add the emission line contributions to the SED templates after verifying an improvement in the comparison between the spectroscopic and photometric redshifts, also according to the parameters $\eta$ and $\mathrm{\sigma_{\Delta z/(1+z_s)}}$.

\begin{figure}
\includegraphics[width=\hsize]{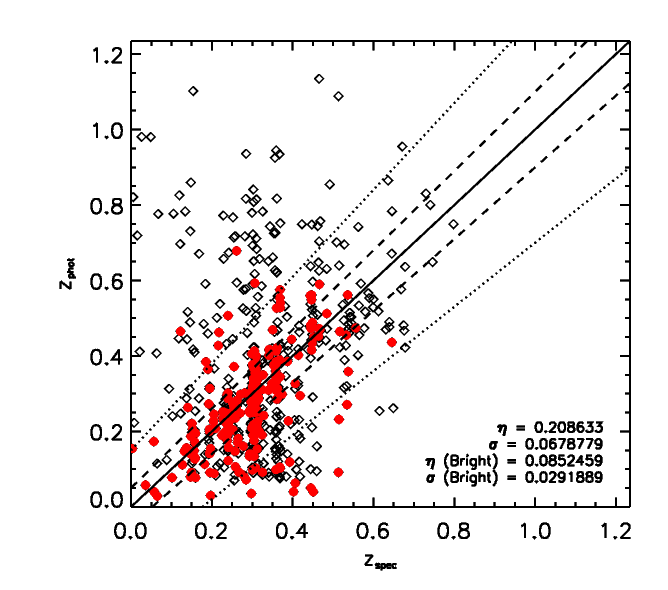}
 \caption{Comparison of spectroscopic and photometric redshifts. Red dots represent all galaxies brighter than $\mathrm{R^\star +1}$. The continuous line is for $\mathrm{z_p=z_s}$, dashed and dotted lines are for $\mathrm{z_p=z_s\pm0.05(1+z_s)}$ and $\mathrm{z_p=z_s\pm0.15(1+z_s)}$, respectively.}
 \label{z_comparison}
\end{figure}
In Fig.~\ref{z_comparison} we show the comparison between $\mathrm{z_s}$ and $\mathrm{z_p}$. The continuous line represents $\mathrm{z_p=z_s}$, the two sets of dashed and dotted lines are for $z_p=z_s\pm0.05(1+z_s)$ and $z_p=z_s\pm0.15(1+z_s)$, respectively. 
The discrepancies in the comparison between $\mathrm{z_s}$ and $\mathrm{z_p}$ is due partly to the small number (i.e. 3) of bands used for the fit and partly to the fraction of catastrophic failures provoked by the misinterpretation of some features. \cite{2007A&A...470..425B} applied their photometric redshift analysis to a simulated catalogue in order to check the robustness of their $\mathrm{z_p}$ against the contamination of high-z outliers. They found that wrong identifications mainly lay outside the cluster photometric range ( $\mathrm{ z\sim 0.3}$).
Thus, after a visual inspection of the outliers in our catalogue at $\mathrm{z \sim 0.3}$ we noticed that the best fit SED templates were typical of late type galaxies, for which the identification of the 4000~$\mathrm{ \AA}$ break is more problematic. Consistently, all these galaxies were classified as faint objects in the VIMOS program (section \ref{spectro_data_descr}) and their spectra revealed some line emissions.

Photometric cluster membership was defined from the distribution of photometric redshifts of spectroscopically confirmed members: we used the biweight mean \citep{1990AJ....100...32B} of their photo-z in order to define the centre of the cluster and a scatter of $\mathrm{\pm 1 \sigma=0.06}$. The redshift interval of the selected cluster members is [0.23,0.35] and the mean $\mathrm{z_{phot}}$ is at $\mathrm{z=0.29}$. Choosing this interval we had a contamination (considering the whole field of the cluster) of 42\% (of which 26\% in foreground and 16\% in background) and we lost 35\% of the spectroscopic sources (of which 18\%  had lower photometric redshifts and 17\% higher ones with respect to the spectroscopic values). For the bright sources (marked in Fig.~\ref{z_comparison} as red dots) we had a contamination of 35\% (of which 25\% in foreground and 10\% in background) and we measured losses of 18\% of the spectroscopic sources (of which 14\%  had lower photometric redshifts and 4\% higher ones with respect to the spectroscopic values). Spectroscopic redshifts were used instead of photometric ones, when available. 
Based on these values, we select a total of 4462 photometric cluster members, 2108 of which are classified as red galaxies and the remaining 2354 as blue, based on their position in the CMD with respect to the red sequence fit given by equation \ref{CMD_equation}.

\subsection{X-ray imaging}
\begin{figure*}
 \includegraphics[width=0.50\hsize]{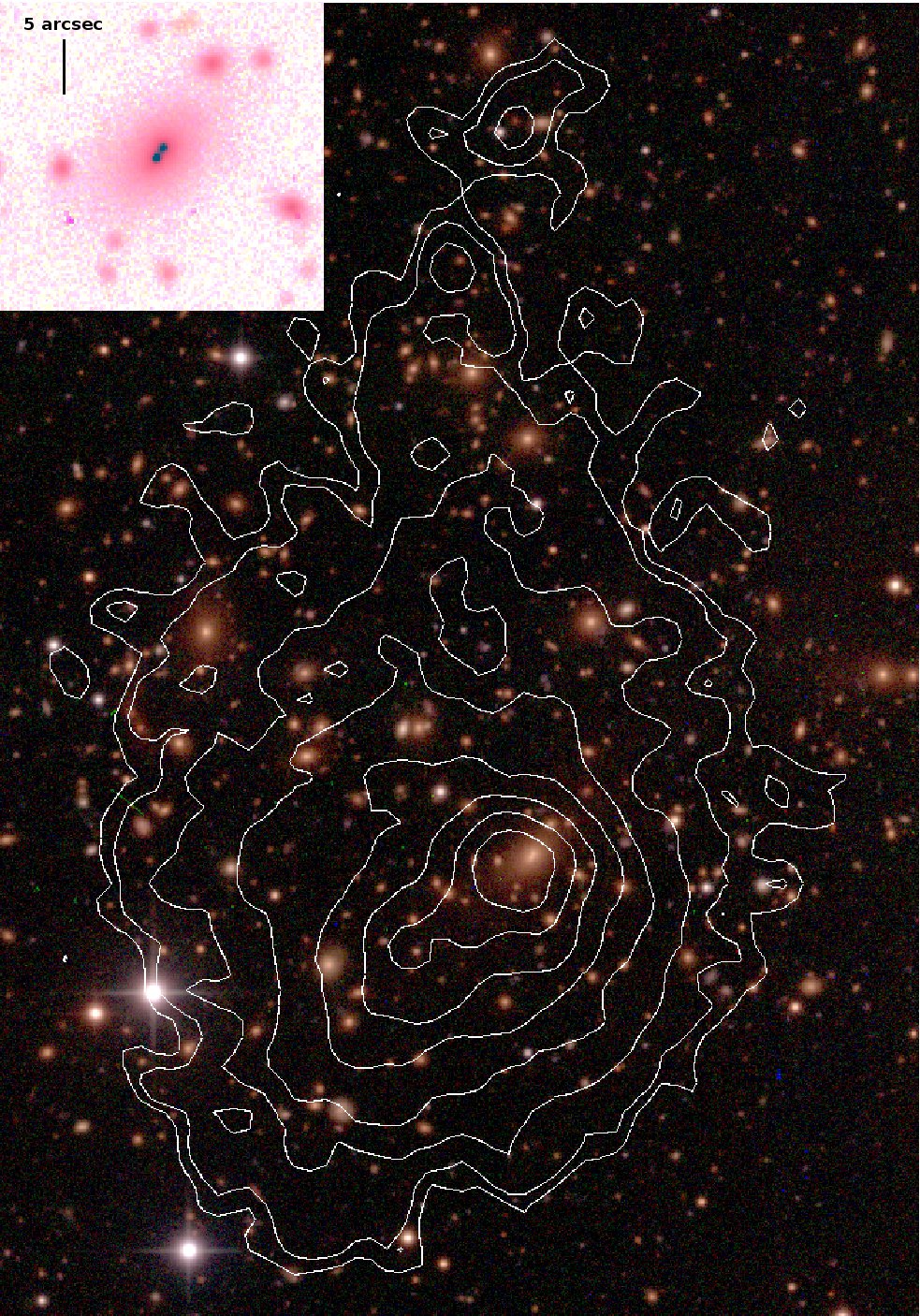}
 \includegraphics[width=0.495\hsize]{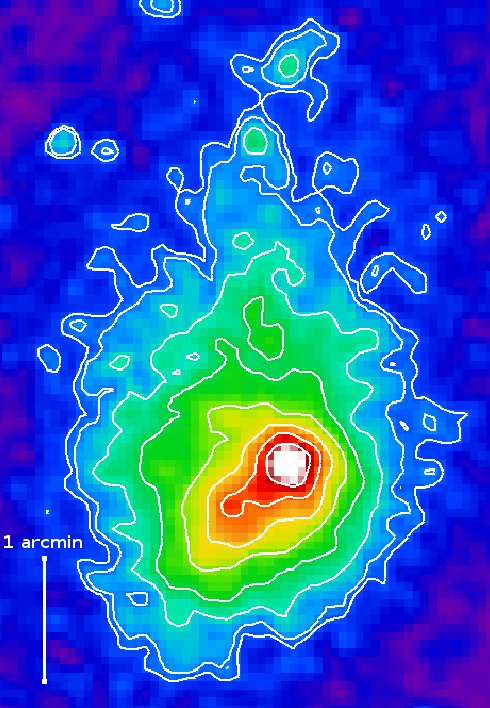}
\caption{Left panel: BVR image of the central region of A1300 with an overlaid composite X-ray emission contour map. Right panel: composite X-ray emission map with contours. The upper left inset shows the Brightest Cluster Galaxy (BCG) for which different cuts and different scales were used for the three bands, in order to highlight the extended halo and the two nuclei separated by about $1.2\arcsec$. On the right: XMM-Newton image of A1300 in the 0.5-2 keV and smoothed with a Gaussian of $\mathrm{\sigma=4\arcsec}$. Z06 report a suspicious X-ray point source at the position (11:31:54.6, -19:55:43), which corresponds to the position of the BCG}.
 \label{BCG_fig}
\end{figure*}

The ROSAT X-ray emission from A1300 was first investigated by \citet{1997A&A...326...34L} who noticed a displacement between the soft (i.e. 0.1-0.4 keV) and hard (i.e. 0.4-2.4 keV) X-ray emission maxima. 
A1300 was observed also by XMM-Newton (Fig.~\ref{BCG_fig}, right panel) in AO-1 as part of the REFLEX-DXL cluster sample in July 2001 (Z06). The total exposure time was $\mathrm{8.8\ ks}$ for EPN and about 14 ks for each of the EMOS detectors.
The data were subjected at this stage to a solar flare cleaning process and the observation was found to be quite clean from contamination so that almost all exposure time survived this process. In fact, after data cleaning from solar flare events we had 8.8 ks for the EPN and 12 ks for the EMOS detectors.

The EPN data were corrected for the out-of-time events in the usual way.
For a more detailed description of the XMM-Newton data reduction see Z06. 
The most important global cluster properties resulting from the XMM-Newton data analysis are the bolometric X-ray luminosity $L=\mathrm{1.80 (\pm 0.15) \times 10^{45}\ erg\ s^{-1} }$, the ICM temperature $T=\mathrm{9.2\pm 0.4\ keV}$ and the mass $\mathrm{M_{500}=5.2 (\pm 3.0) \times 10^{14} M_{\sun}}$  (Z06).

The right panel of Fig.~\ref{BCG_fig} shows an XMM-Newton image of A1300 in the 0.5-2 keV band, smoothed with a Gaussian of $\mathrm{\sigma=4\ \arcsec}$. Z06 report a suspicious X-ray point source at the position (11:31:54.6, -19:55:43), which corresponds to the position of the BCG. The cluster exhibits an elliptical morphology according to the classification of the dynamical state based on X-ray imaging \citep{1992csg..conf...49J}. The X-ray peak is displaced with respect to the BCG by 36 kpc. The left panel of Fig.~\ref{BCG_fig} shows the same X-ray contours of the right panel superimposed on the BVR image of the central part of A1300 whose BCG presents 2 nuclei at its centre (upper left corner of Fig.~\ref{BCG_fig}).

In order to obtain the maximal information from our X-ray data we performed the analysis using the method of the PSF reconstruction as explained in \citet{2009ApJ...704..564F}. In particular, after the point source removal, we computed the background estimate. 
This procedure allowed us to highlight not only the dynamical shape of the cluster, but also to determine the significance of the structures already seen in an appropriately smoothed X-ray image, as better explained in Section \ref{substructure_id_paragraph} and Appendix \ref{substructures_optical}. This enables us to reconstruct the dynamical history of the cluster through the analysis of the groups and their link with the large scale structure of A1300.

\section{Results and Morphological Overview}

\subsection{Cluster morphology and large-scale galaxy distribution}
\label{density_maps_comp_section}
The red sequence fit (Fig.~\ref{CMD} and Equation~\ref{CMD_equation}) allows us to separate red and blue galaxies, for which we derived the galaxy density distribution using an adaptive kernel smoothing. This method is a refinement of the basic procedure discussed in B09 and provides a more accurate estimate of the background counts and noise, while retaining all significant information about substructure and large-scale structure, also in the outskirts of the cluster (i.e. beyond $\mathrm{R}_{200}$).
\begin{figure}
\includegraphics[width=\hsize]{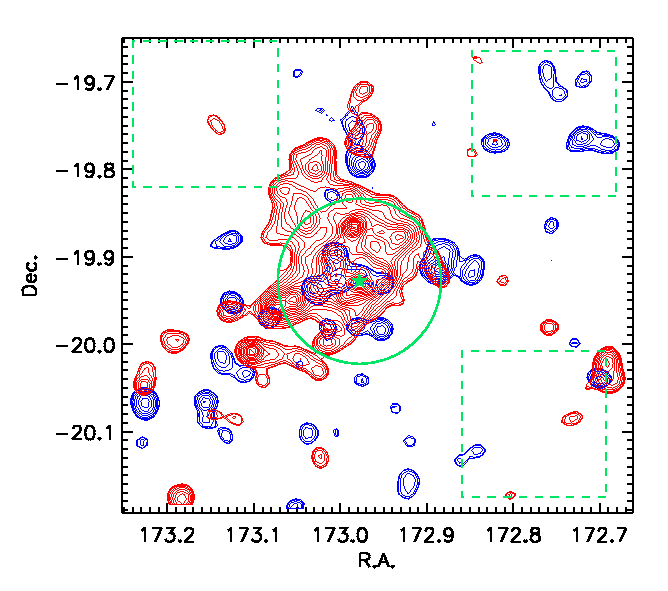}
\caption{Density contours for red and blue photometric members of A1300 smoothed with an adaptive kernel. Red galaxies (in red) are more concentrated in the central regions, while blue galaxies (in blue) are preferentially located in filamentary structures. The green star shows the peak of the X-ray emission,   the region of the cluster within $\mathrm{R_{200}}$ is marked by the green circle, while the three green dashed-lined boxes highlight the regions selected for estimating of the background. The contours have a significance of at least $5 \sigma$ w.r.t. the background and follow a square root scale.}
 \label{density_maps}
\end{figure} 

Red galaxies (red contours in Fig.~\ref{density_maps}) are concentrated towards the innermost region of the cluster with a mean galaxy density of 1.57 galaxies $\mathrm{arcmin^{-2}}$), which is highly significant w.r.t. the background. Conversely blue galaxies (in blue) are more scattered and mostly located in extended structures beyond $\mathrm{R_{200}}$. 

The over-density information from the galaxy distribution can be used to estimate the amount of galaxy mass belonging to the cluster and compare it to that in the filamentary regions in the NE and SE. Using the X-ray peak as a reference centre and  $\mathrm{R_{200}}$ as the fiducial radius of the cluster, we find that the cumulative luminosity of all red galaxies in the NE (SE) filament is about 30\% (33\%) of the total. Such a high fraction of red (i.e. likely passively-evolving) galaxies beyond $\mathrm{R_{200}}$ suggests that the infalling galaxies may have evolved already along the large-scale structure before  falling into A1300. This might have happened in the lower-mass groups through the same filaments.

\begin{figure}
\includegraphics[width=\hsize]{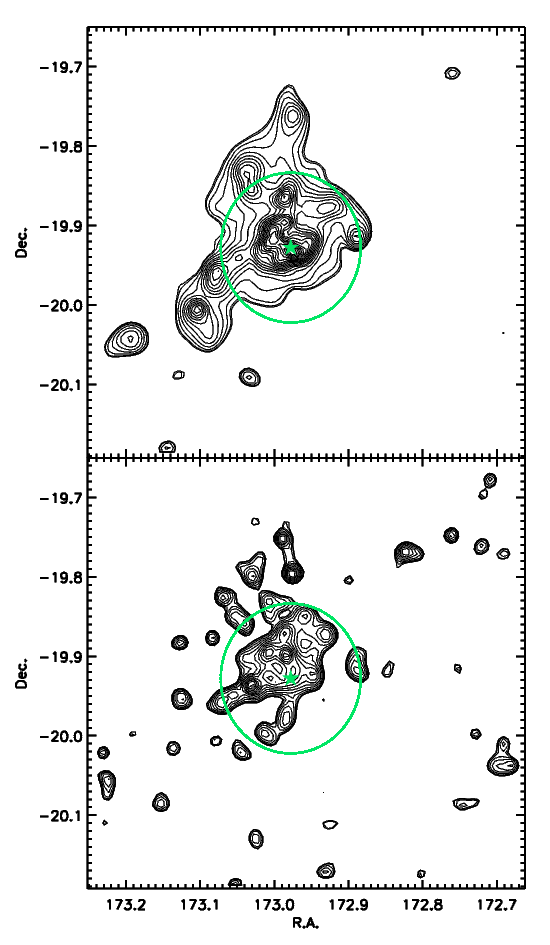}
\caption{Density contours for bright ($\mathrm{R< R^{\star} +1}$, top panel) and faint ($\mathrm{R\geq R^{\star} +1}$,bottom panel) photometric members (both red and blue galaxies) of A1300 using an adaptive kernel method. The green star represents the peak of the X-ray emission while  the region of the cluster within $\mathrm{R_{200}}$ is marked by the green circle. The contours have a significance of at least $5 \sigma$ w.r.t. the background (corresponding to 0.41 galaxies ${\rm arcmin^{-2}}$ for the bright cluster members and 3.32 galaxies ${\rm arcmin^{-2}}$ for the faint ones)  and follow a square root scale.}
 \label{density_map_bright}
\end{figure}

\begin{figure}
\includegraphics[width=\hsize]{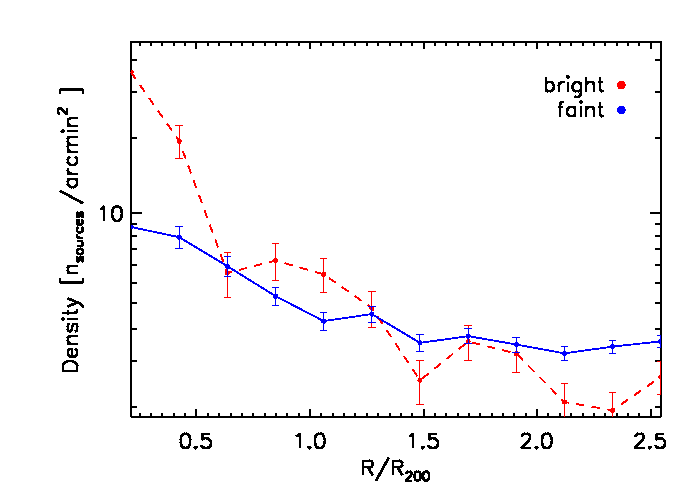}
\caption{ Density profile for bright (red dashed line) and faint (blue solid line) photometric members (both red and blue galaxies) of A1300. The density profile of the bright galaxies is normalised to the same total number of faint galaxies in order to compare it with that of the faint galaxies. Bright objects are mostly located in the central region of the cluster while faint ones dominate in the outskirts.}
\label{density_profiles}
\end{figure}

Further comparing the density maps for all bright ($\mathrm{R< R^{\star} +1}$) and for faint ($\mathrm{R\geq R^{\star} +1}$) galaxies (both red and blue ones, Fig.~\ref{density_map_bright}), two different behaviours can be identified. On one hand, bright galaxies are found at larger ratio close to the X-ray peak (as confirmed by the density profiles in Fig.~\ref{density_profiles}), with some clumps following the overall direction of the filaments, while on the other hand, faint galaxies (present in larger numbers) show more over-dense clumps all across the cluster and its outskirts. Hence it seems that, while massive galaxies trace better the inner region of the cluster and the surrounding large-scale structure, faint galaxies  dominate in the outskirts (at $\rm R \gtrsim 1.3\ R_{200}$) and can provide insight to the substructure at the smaller scales.

\subsection{Temperature, Pressure and Entropy}
\label{temp_entr_section}
The X-ray properties for the DXL sample were first investigated in \citet{2004A&A...413...49Z}. A1300 exhibits a temperature gradient and cool intra-cluster gas in the centre, suggesting that cooling cores are not only found in clusters with symmetric and regular X-ray images, but can also be found in elongated, very disturbed clusters (such as A1300). An estimate of the cooling time was given in  Z06: $\mathrm{t_{cool}\approx10\ Gyr}$, assuming a gas temperature $\mathrm{T=10^8\ K}$ and a density $\mathrm{n_p= 5.8 \times 10^{-3}\ cm^{-3}}$. The region where the cooling time is smaller than the age of the Universe at the cluster redshift was found to be within a cluster-centric distance of  $27\arcsec$ ($\sim$120 kpc) at maximum.

\begin{figure*}
 \includegraphics[width=0.48\hsize]{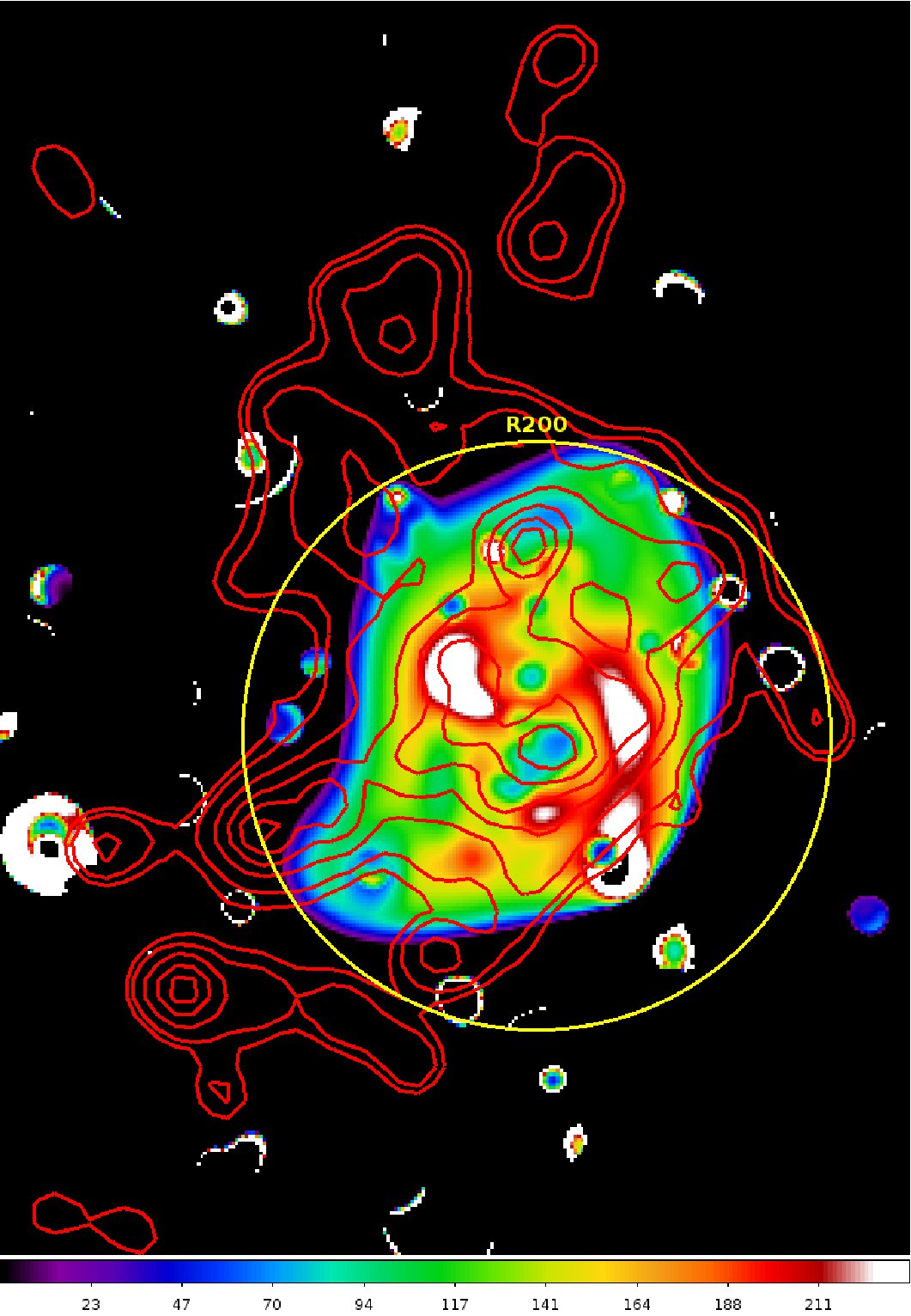}
 \includegraphics[width=0.48\hsize]{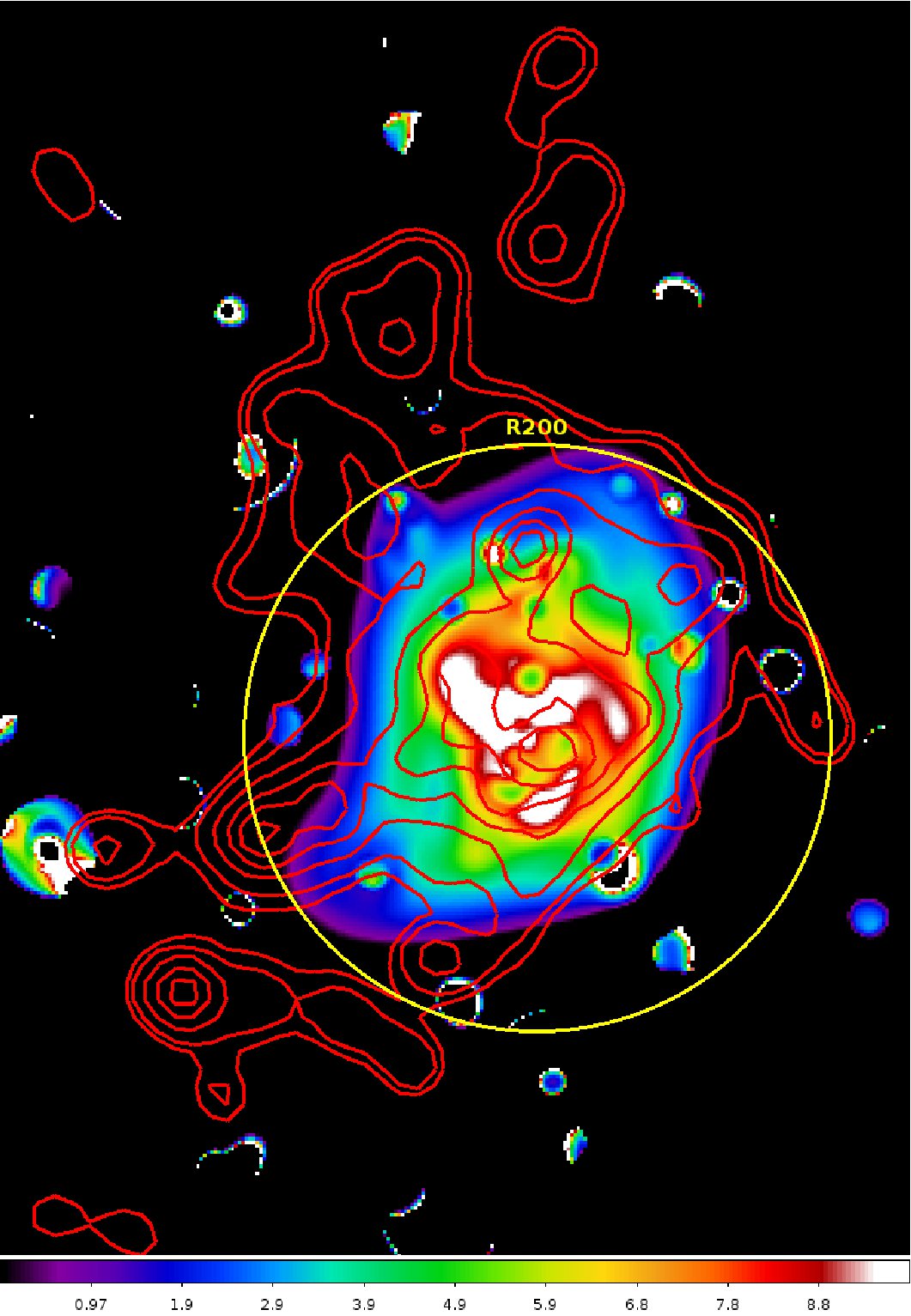}
\caption{ On the left: Entropy pseudo-map of A1300. The colour bar is in arbitrary units. On the right: Temperature map of A1300. The colour bar is in keV (For details on both pseudo-maps see \citealt{2005A&A...442..827F}). In both panels red contours outline the over-density of red galaxies while the yellow circle represents the region of the cluster within $\mathrm{R_{200}}$. }
 \label{entropy_map}
\end{figure*}

A more qualitative analysis of the temperature, pressure and entropy\footnote{The entropy is an important diagnostic parameter because it determines the structure of the ICM recording its thermodynamic history. We adopt the definition of \citet{2005RvMP...77..207V} for the entropy: $\mathrm{K= k_b T n_e^{-2/3}}$, where $\mathrm{k_b}$ is the Boltzmann constant, T is the temperature in keV units and $\mathrm{n_e}$ is the electron density.} maps was performed by \citet[we address the reader to their Fig.~11]{2005A&A...442..827F} who named A1300 the ``Whirlpool'' cluster of galaxies because of the features in its temperature map.

\citet{2005A&A...442..827F} found a central East-West ridge of high temperature that may reflect the compression of the central region between the two main merging components.
A distorted cool core, which partially preserves the characteristic low entropy, could be responsible for the complex temperature structure of the cluster. Moreover, the same authors found a large scatter in the entropy profile arguing that it reflects a high degree of substructure in the cluster. 

Therefore, we compare the spatial distribution of galaxies and hot gas in order to investigate consistent behaviours of both components. In particular, the entropy map ( left panel of Fig.~\ref{entropy_map}) provides a reliable record of the gas history \citep{2005RvMP...77..207V}. The distribution of the member galaxies (red in particular) follows closely the entropy features, suggesting that galaxies track the information provided by the gas in the central regions out to larger cluster-centric radii.
A similar behaviour can be seen in the temperature map (right panel of Fig.~\ref{entropy_map}) where the hot features are probably caused by shocks resulted from the collisions of different merging components. 
This comparison reveals a direct correspondence between the substructures traced by the gas and the galaxies.
 After the impact of two massive clusters it is likely that the coupled gas and dark matter components of each cluster started to swing around a common centre yielding the characteristic whirlpool shape seen in Fig.~\ref{entropy_map}. The projection effects play an important role as, to our knowledge, there is no other cluster with a spiral like shape of the galaxy density distribution which resembles the gas temperature or entropy map.
The merging process could also have affected the star formation activity of the galaxies. The interaction with the ICM could have provoked the quenching of the star formation, possibly, after a starburst (e.g. \citealt{2004ApJ...601..197P}). 
Taking this topic is far from the aim of this paper as further investigation will rely on the ongoing spectral index analysis (Ziparo et al. in preparation).

\subsection{X-ray Surface Brightness and Identification of Substructures}
\label{substructure_id_paragraph}
\begin{figure}
\includegraphics[width=\hsize]{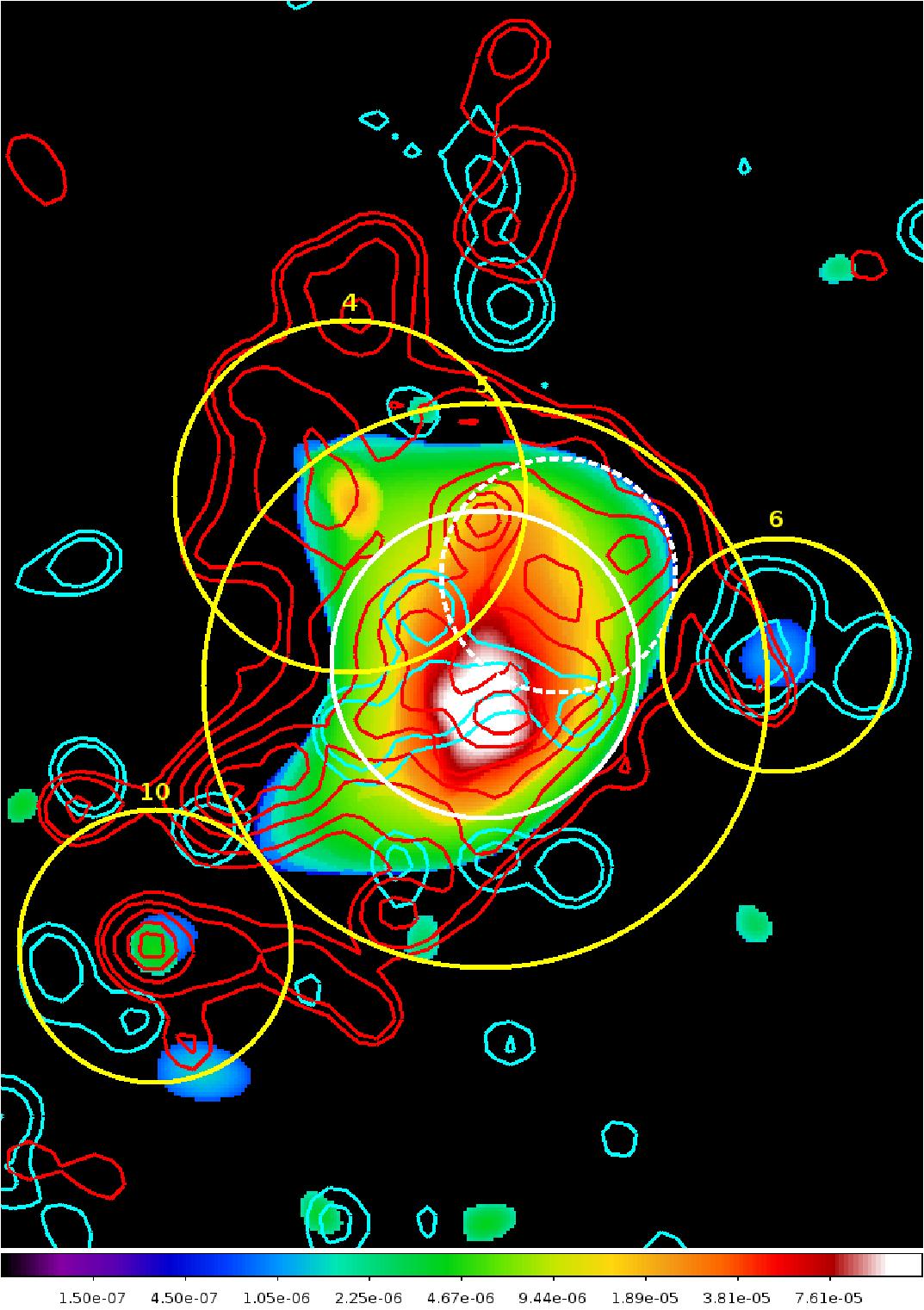}
\caption{Surface brightness in the $\mathrm{0.5-2\ keV}$ band of A1300 using the technique described in \citet{2009ApJ...704..564F}. Several extended sources are detected around the cluster, with a significance larger than 4$\sigma$ w.r.t. the background. Superimposed are the density contours of red and blue cluster members. The yellow circles represent $\mathrm{R_{200}}$ of each group identified with the same ID of Table \ref{groups_table} (number 5 is A1300). The continuous white circle shows how the X-ray emission of the cluster would look like if it were spherically symmetric and the dashed white circle a possible group that disturbs the original symmetry but is now dissolved inside the cluster. The units of the colour bar are in counts/sec/pixel. }
 \label{psf_rec_dens_maps}
\end{figure}

In order to minimize the impact of point sources and isolate the X-ray emission due to the diffuse hot ICM, we apply a novel technique  to enhance the significance of extended sources and filter out the point sources from the X-ray surface brightness map \citep{2009ApJ...704..564F}. Although this method was originally designed for group identification in wide-field/survey areas, it proved to be also quite efficient in identifying and characterising smaller subsystems within or close to clusters. In particular, where point sources were detected in the XMM-{\it Newton} maps (and confirmed by archival {\it Chandra} data), this technique enabled us to reduce their X-ray appearance and better remove their contribution from the extended emission (for further details and explanations see \citealt{2009ApJ...704..564F}). 

Figure \ref{psf_rec_dens_maps} shows the X-ray surface brightness in the $\mathrm{0.5-2\ keV}$ band obtained using the wavelet+PSF reconstruction. It is possible to identify several extended sources in the proximity of the cluster (with a significance larger than 4$\sigma$ w.r.t. the background in the optical). However the X-ray emission itself is not enough to establish their membership to the A1300 system.

\begin{figure}
\includegraphics[width=\hsize]{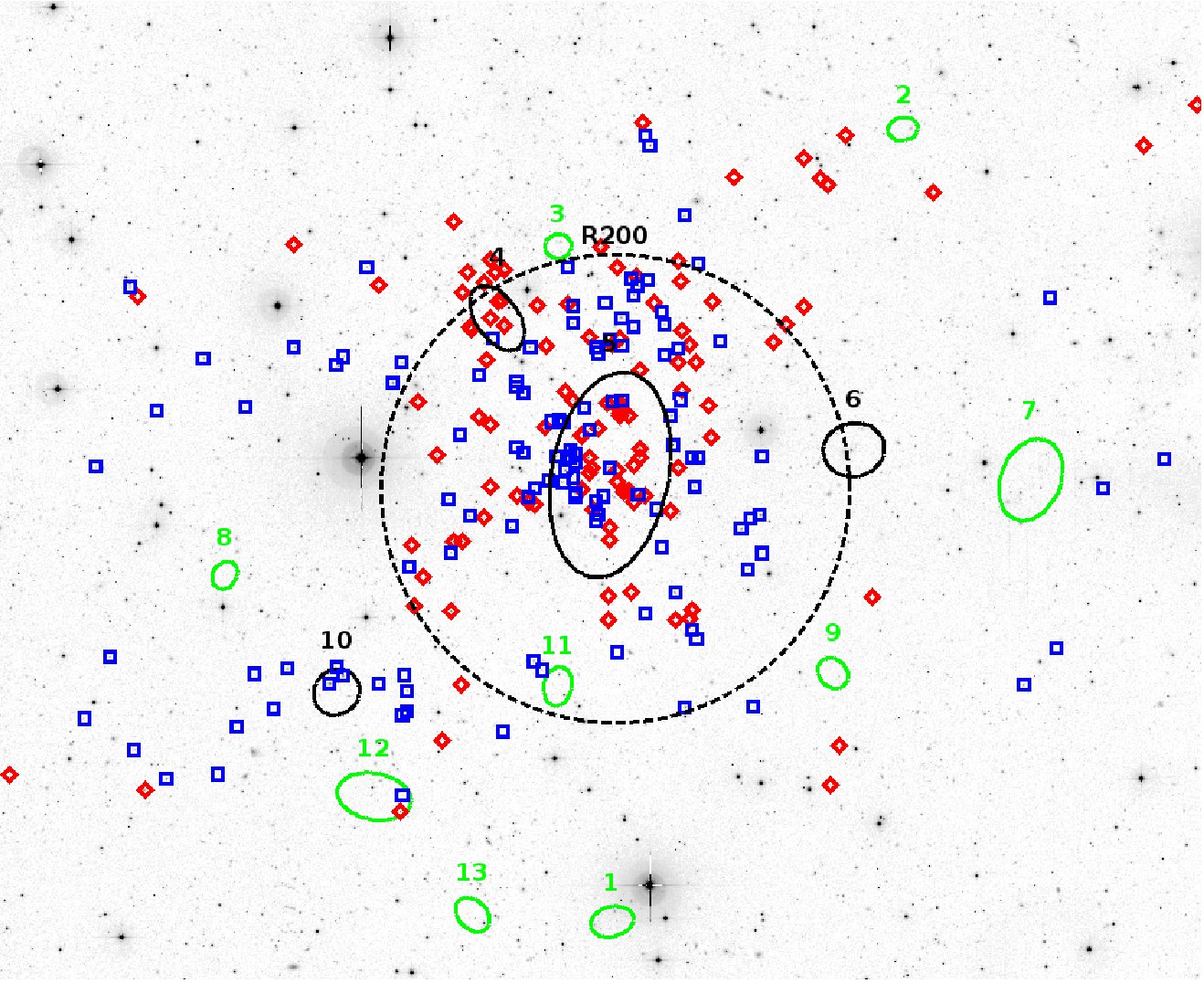}
\caption{WFI R band image of A1300. The ellipses represent the groups observed in this field: the shape of all groups is related to their X-ray emission (Fig.~\ref{psf_rec_dens_maps}). Black ellipse identify groups with a high probability to belong to the cluster  i.e. at a significance larger than 5$\sigma$ w.r.t. the background, the green ones have a lower probability. Group number 5 is the main cluster of which region inside $\mathrm{R_{200}}$ is represented by the black dashed circle. The small symbols are the spectroscopic member galaxies of the cluster: the blue squares are all the galaxies approaching with respect to the observer, the red diamonds are the receding ones. Group 4 and 10 correspond to the substructures found in the DS-test (red circles in Fig.~\ref{dstest}).}
 \label{groups_appr_rec}
\end{figure}

Comparison with the distribution of cluster members shows that a few extended X-ray sources match the position of over-densities in the galaxy distribution maps. The results of this combined analysis and identification of the groups in X-rays and optical are described extensively in Appendix \ref{substructures_optical}.
All groups detected and possibly related to the cluster are marked in Fig.~\ref{groups_appr_rec}: in black are those belonging to the LSS of A1300 (detected at more than 5$\sigma$ w.r.t. the background), whereas those in green have a lower probability to be at the same redshift of the cluster.
The groups selected as related to the cluster (ID numbers 4, 6 and 10) are also highlighted by yellow circles in Fig.~\ref{psf_rec_dens_maps}. They all lie in the outskirts of the cluster, thus providing a tool to probe the accretion regions and to investigate the large-scale dynamics of A1300.

The central region of the cluster (marked by the solid white circle in Fig.~\ref{psf_rec_dens_maps}) is spherically symmetric. This region has a radius of 3.09 arcmin (corresponding to about $\mathrm{835\ kpc}$ at the cluster redshift) and emphasizes a possible forward shock towards SW, followed closely by the galaxy density distribution.
This shock is possibly related to the candidate radio relic found in previous studies (\citealt{1999MNRAS.302..571R} and \citealt{2011arXiv1102.1901G}).

\begin{figure}
\includegraphics[width=\hsize]{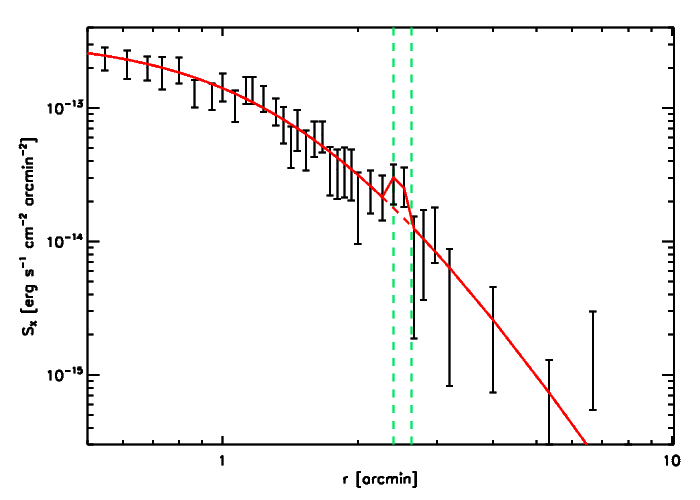}
\caption{0.5-2 keV surface brightness profile of A1300 extracted from the centre of the curvature of the shock towards the SW quadrant, where the candidate radio relic was found (\citealt{1999MNRAS.302..571R}, \citealt{2011arXiv1102.1901G}) at the position (11:31:46.8, -19:56:44). The errors are 1$\sigma$, the red continuous line shows the best fit model while the dashed line shows the same model in the region of the shock without the enhancement. The green dashed lines constrain the position of the relic which corresponds to the enhancement we find in the profile. }
 \label{shock}
\end{figure}

Fig.~\ref{shock} shows the 0.5-2 keV surface density radial profile extracted in the SW quadrant in the direction of the radio relic (the position of which is highlighted by the green dashed lines). We used the position of the relic to tentatively identify the curvature of the shock (as already done, for example, by \citealt{2011ApJ...728...82M}). We find an enhancement in the surface brightness coincident with the position of the relic. We use only pn data in which this enhancement is more evident after masking all point sources and extended emissions outside the cluster. Our best fit model is obtained with a projected emissivity profile in which the density jump in the position of the relic is ${\rm \rho_1/\rho_0 = 1.30 \pm 0.15}$, where $\rho_0$ is the density of the unperturbed gas and $\rho_1$ is the density of the gas after the shock.  
The density jump allows us to derive an upper limit to the strength of this possible shock. Assuming a monoatomic gas with ${\rm \gamma=5/3}$ and using equation 1 of \cite{2010ApJ...715.1143F} we obtain a Mach number $\mathcal{M}=1.20 \pm 0.10$.

\cite{2005A&A...442..827F} define the cluster merging direction with the North-South on the basis of their X-ray qualitative analysis. Using X-ray and optical information (thus through a more detailed study) we are able to better define this direction. 
However the study of the dynamics of this cluster is not straightforward.

 In fact, the X-ray surface brightness map shows a more complex morphology in the outer regions of the cluster, with multiple extended structures around and beyond $\mathrm{R_{200}}$ and a globally asymmetric shape. This matches the distribution of photometric cluster members (cf. Fig.~\ref{psf_rec_dens_maps}), suggesting that  A1300 is accreting matter along filamentary structures. In particular, the extended structure seen to the SE of the main cluster in the distribution of galaxies matches the position of Group 10, suggesting that this system may be infalling along a filament which extends from SE to NW.

In addition, we highlight with a dashed white circle in Fig.~\ref{psf_rec_dens_maps} a possible group which is now part of the cluster: its identity is lost in X-rays but it is visible in the optical image as a concentration of galaxies and in Fig.~\ref{psf_rec_dens_maps} as a density peak.
To the NE of the core, the symmetry is disturbed by group 4 (shown in Fig.~\ref{group_snapshots_and_CMD} with overlaid X-ray contours) which lies mostly inside the $\mathrm{R_{200}}$ of the cluster. Even if this group is very close to the cluster, it preserved its identity both in the X-ray and in the optical: its X-ray emission stands out even if it is already embedded with that of the cluster  (Fig.~\ref{psf_rec_dens_maps}). It is also possible to trace a red sequence (Fig.~\ref{group_snapshots_and_CMD}) and to identify a BCG in the corresponding region in the optical image. The X-ray mass of this group is  $M_{200}=\mathrm{1.35\times10^{14}\ M_{\sun}}$ (obtained adopting the same scaling relations of \citealt{2010ApJ...709...97L}, i.e. after assuming a beta profile and removing embedded point sources, see Appendix~\ref{substructures_optical} for details), i.e. 10\% of that of the cluster. 
We are thus able to reconstruct the accretion pattern of A1300, which looks to be entering a phase of dynamical relaxation in its inner region, while still accreting mass in its outskirts.

\subsubsection{Substructures from kinematics}

We apply the kinematical DS-test \citep{1988AJ.....95..985D} to the dataset of available spectroscopic cluster members, to identify substructures based on the combination of their position and velocity. This provides a complementary approach to detecting over-densities in the galaxy distribution and has the potential to identify substructures bound to the cluster but still retaining their kinematical identity.

\begin{figure}
\includegraphics[width=\hsize]{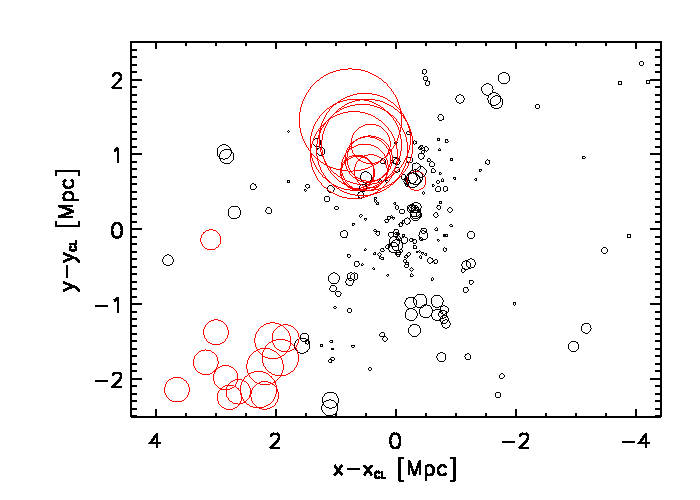}
\caption{Substructure skyplot showing the results of the DS-test for the spectroscopic members of A1300. Circles are centred on the positions of member galaxies; their radius is proportional to $e^{P_{DS}}$, where $P_{DS}$ is the DS deviation parameter for each galaxy. Red circles mark groups of galaxies which exhibit significant deviations ($>3\sigma$) from the local velocity distribution.}
\label{dstest}
\end{figure}

The test was iterated over $10^4$ re-samplings of the velocities and yields a DS statistics $P_{DS} = 4\times10^{-4}$, implying that the cluster has a significant degree of sub-structure (the whole test being significant at $5.5\sigma$). The $\kappa$-test of \citet{1996ApJ...458..435C} was also run for comparison on the same dataset with consistent results. The results of the DS-test are shown in Fig.~\ref{dstest}.

We identify kinematical substructures as groups of galaxies with values of the DS parameter beyond a critical value of 2.58 (calculated from the re-sampling statistics). Two groups, both with a significance above the $3\sigma$ level, were detected.

A first group to the NE of the cluster centre consists of 12 cluster members with a mean peculiar velocity of +1494$\,\rm{km~s^{-1}}$ w.r.t the mean cluster redshift and a velocity dispersion of 582$\,\rm{km~s^{-1}}$. This group appears quite compact on the plane of the sky, and its mean position (11:32:09.4, -19:50:43.8) is consistent with that found in X-rays for group 4. All 12 galaxies lie within the $\rm{R_{200}}$ of this group.
After correcting for velocity errors, we derive a tentative mass estimate from the velocity dispersion (using the relation given by \citealt{2006A&A...456...23B}), finding a mass of $\mathrm{(1.17 \pm 0.24)\times10^{14}\ M_{\sun}}$, in good agreement with the X-ray estimate (cf. Appendix \ref{substructures_optical}).

Another group of 7 galaxies to the SE of the cluster centre exhibits large systematic deviations. These galaxies appear less concentrated than those identifying the previous group, however they all lie within $R_{200}$ of group 10, their mean position (11:33:02.8, -20:12:09.33) being only 53$\arcsec$ away from the X-ray peak (Table \ref{groups_table} and ellipse number 10 in Fig.~\ref{groups_appr_rec}). We find a mean velocity of -774$\,\rm{km~s^{-1}}$ w.r.t the cluster mean velocity and a dispersion of 452$\,\rm{km~s^{-1}}$, from which we infer a mass of $\mathrm{(4.37 \pm 2.99)\times10^{13}\ M_{\sun}}$. Although with larger uncertainty, also for this group we find an agreement with the X-ray value.

\subsection{Linking merging configuration and substructures}

The combination of the X-ray, photometric and spectroscopic data enables us to investigate the dynamics of the cluster and to link them to its main substructures.

Figure \ref{groups_appr_rec} shows all spectroscopic cluster members grouped by their peculiar velocity w.r.t the cluster mean velocity: receding from the observer in red, approaching in blue. Ellipses mark the X-ray groups detected with the PSF reconstruction technique, groups encircled in black being those with a high probability to belong to the cluster. The cluster $\mathrm{R_{200}}$, marked with the black dashed circle, highlights the presence of three groups (ID 4, 6, 10) entering the virialised region of the cluster.

The dynamical configuration of A1300 is quite complex, as already anticipated in the previous sections. X-ray and entropy maps reveal signs of a major merger and the presence of three X-ray groups. Two of these (ID 4 and 10) find confirmation from both the distribution of galaxies and the kinematical substructures and they seem to be accreted onto the main cluster,  while a third group (ID 6) may be following a third accretion direction in a filament pointing towards the observer (unfortunately we do not have spectroscopic information to confirm it).

Northwards of the main cluster, Group 4 shows a significantly higher recession velocity by +1494$\,\rm{km~s^{-1}}$. Our findings suggest that this group may be part of the northern filament; from its peculiar velocity  we can infer that the filament lies between the observer and the cluster. This is mirrored by the global velocity distribution of galaxies to the North of the cluster, which shows a systematic velocity of +456$\,\rm{km~s^{-1}}$ for galaxies outside the core ($\sim$1 Mpc away from the cluster centre).

Towards the South, a filamentary structure is traced by photometric spectroscopic cluster members. Group 10 is detected in the same region. It is particularly significant in X-rays and in the optical as well, showing a rich red sequence (see Fig.~\ref{group_snapshots_and_CMD}) and a well-defined over-density of galaxies (see Tab.~\ref{groups_table}). It is also detected from kinematics, with a peculiar velocity of -774$\,\rm{km~s^{-1}}$. Consistently, the overall velocity field in the southern outskirts is negative, with a mean of -354$\,\rm{km~s^{-1}}$. This suggests that the filament in which Group 10 is embedded reaches the cluster from behind.

\begin{figure}
 \includegraphics[width=\hsize]{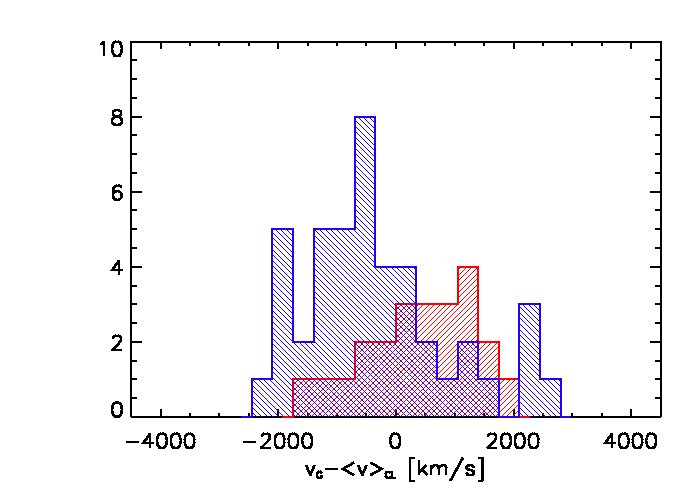}
 \caption{Velocity distribution for galaxies further than 1 Mpc from the cluster centre. Red histogram: northern outskirts; blue histogram: southern outskirts.}
 \label{histo_vel_north_south}
\end{figure}
The velocity distributions in the two regions (1 Mpc northwards of the cluster core and 1 Mpc southwards of it) are shown in Fig.~\ref{histo_vel_north_south}. We compare the two distributions by means of a two-sided Kolmogorov-Smirnov test: this shows that the two distributions are significantly different, yielding a KS parameter of 0.018 (i.e. a significance of $3.3\sigma$).

We can estimate the expected infall velocity around $\mathrm{R_{200}}$ for a group infalling from a large distance. Assuming a mass of $10^{15} \mathrm{M_{\sun}}$ for the main cluster as calculated for system 5 (cf. Appendix A), we obtain a velocity of about $\mathrm{2000\ km/s}$ at a distance of ${\rm 2\ Mpc}$ from the cluster. This allows us to estimate for each filament the angle w.r.t the plane of the sky: we find that the northern filament has an angle of about 40$^{\circ}$, while the southern has a lower angle of about 20$^{\circ}$. In addition, given the estimates of total luminosity for the two filaments (cf. Section \ref{density_maps_comp_section}), we estimate that A1300 will accrete about 60\% of its current total mass the next Gyr from infalling material.

\section{Discussion}
A1300 is a post-merging cluster at $\mathrm{z\sim 0.3}$ with a ``dumbbell'' cD galaxy at its centre and prominent filaments visible in the galaxy density distribution.
The BCG position is not coincident with the position of the X-ray emission peak, on the other hand, this cluster exhibits an elongated X-ray emission and a disturbed galaxy distribution in the optical.
As already found by B09 in other two DXL clusters (and in the past in other clusters e.g. by \citealt{1980ApJ...236..351D}), also A1300 hosts a centrally located population of galaxies dominated by an old, passively evolving stellar populations which define the red sequence. 

The study of the CMD enables us to measure the magnitude gap between the BCG and the second brightest galaxy. We can tentatively use it as a proxy on the dynamical state of the clusters (as already done in greater detail for a sample of galaxy clusters by \citealt{2010MNRAS.409..169S}). 
Table \ref{mag_gap_table} provides a comparison of the gap statistics for the REFLEX-DXL clusters studied so far.
RXCJ0014.3-3022, which is in an early phase of merging, displays the smallest magnitude gap. 
A1300 (RXCJ1131.9-1955), classified as a post-merging system \citep{1997A&A...326...34L}, has an intermediate value of $\mathrm{\Delta m}$, while RXCJ2308.3-0211, which is a very regular cluster with a cool core, has the largest gap. This supports the conclusions by \citet{2010MNRAS.409..169S} i.e. the time passed since the last major merger is directly reflected in the growth of the BCG; in this way the magnitude gap increases with the ageing of the cluster.

\begin{table}
\centering
\resizebox{\hsize}{!}{
\begin{tabular}[!t]{ccccc}
\hline

RXCJ & a & b & $\mathrm{\Delta mag}$ & Dynamical State \\ 
\hline

0014.3 \textendash 3022  &   $-0.037 \pm 0.003 $ & $2.935 \pm 0.238$ & $\sim 0.1$ & Merging \\ 
1131.9 \textendash 1955  &   $-0.048 \pm 0.005 $ & $2.823 \pm 0.090$ & $\sim 0.8$ & Post-merging \\ 
2308.3 \textendash 0211  &   $-0.038 \pm 0.003 $ & $2.933 \pm 0.323$ & $\sim 1.1$ & Cool core \\
\hline

\end{tabular}
}
\caption{Optical properties for 3 DXL clusters: col.1 gives the name of the cluster as in the REFLEX catalogue, col.2 and col.3 the fit parameters of the best red sequence fit (where the fit is represented by the equation $\mathrm{y=ax+b}$), col.4 the gap between the two brightest galaxies and col.5 the dynamical state.}
\label{mag_gap_table}
\end{table}

 Table \ref{mag_gap_table} also provides information on the parameters of the best fitting line to the red sequence for the 3 DXL clusters.

The red sequence fit in the CMD (Fig.~\ref{CMD}) allows us also to separate red and blue galaxies, for which we derive the galaxy density distribution (Fig.~\ref{density_maps}).
The density contours of red galaxies reveal a vortex-like shape reminiscent of the temperature and the entropy features (Fig.~\ref{entropy_map}). The filamentary structures are highlighted mostly by the red galaxies departing from the central region of the cluster and extending beyond $\mathrm{R_{200}}$. This suggests that infalling galaxies have already evolved before starting their infall towards A1300, possibly in former groups embedded in the surrounding feeding filaments. In fact, the gravitational
potential well of the groups could affect the evolution of their galaxies \citep{2004ogci.conf..426L}. Furthermore, the extent to which galaxies are pre-processed in groups before falling into clusters depends on the mass of galaxies \citep{2009MNRAS.400..937M}, suggesting that massive galaxies are more easily segregated in the group halos.

A possible shock front (consistent with a Mach number $\mathcal{M}=1.20 \pm 0.10$) was also identified in the southern part of the cluster coincident with a candidate relic (\citealt{1999MNRAS.302..571R} and \citealt{2011arXiv1102.1901G}) found in the radio bands. 
Radio halos are diffuse radio emission located at the centre of clusters, while relics are usually elongated or arc-shaped and found in the peripheral regions of clusters (e.g. \citealt{2011arXiv1102.1572V} for a recent review).
Most of the merging clusters host both radio halo and relics \citep{2002ASSL..272..197G}, likely originated by acceleration of electrons through shocks and turbulence \citep{2011arXiv1106.0591B}.  
The Mach Number that we find from the surface brightness profile is consistent with a shock originating from a merging of two progenitors with comparable mass \citep{2007PhR...443....1M}. The gas motion after the merging could have played an important role though.

Surprisingly also the galaxy distribution exhibits a sharp edge in the SW. This coupled conduct of gas and galaxies is also visible in the analysis of the temperature and entropy maps. In particular, we overlaid the density contours of the red galaxies of the cluster (mostly populating the inner region) which nicely follow the entropy (and temperature) features, suggesting that the galaxies track the information provided by the gas in the central region out to larger cluster-centric radii. The projection effects combined with the status of the merging play an important role as, to our knowledge, there is no other cluster with a spiral like shape and the galaxy density distribution which reminds the gas temperature or entropy map.

\citet{2011ApJ...728...54Z} performed simulations of cluster merging starting with two cool core clusters with different masses and total mass distributions (gas and dark matter) represented by a Navarro, Frenk and White (NFW, \citealt{1997ApJ...490..493N}) profile. The status of A1300 seems to be represented by one of the simulations where two parent clusters of approximately the same mass ($\mathrm{M_{200}\sim 6 \times 10^{14}\ M_{\odot}}$) are approaching one another with an initial impact parameter $\mathrm{b\approx 464\ kpc}$  (fig.~4 for simulation S2 in \citealt{2011ApJ...728...54Z}). However, we can not exclude the possibility represented by the same simulation with an impact parameter $\mathrm{b\approx 932\ kpc}$ (fig.~5 for simulation S3 in \citealt{2011ApJ...728...54Z}). 
In these simulation the gas cores do not collide at the first core passage but sideswipe, creating a bridge of stripped, low entropy gas that stretches between the two DM cores. In fact, as high entropy gas floats and low-entropy gas sinks, the low entropy region marks the positions of the core of the cluster progenitors and the high entropy is due to the presence of shocks. 
The entropy map of the simulation finds a pretty good correspondence to the projected entropy map of A1300: here one plume of cold gas is located in correspondence of the BCG, showing up as a low entropy channel in the central region of the cluster, surrounded by higher entropy gas (with a characteristic whirlpool shape).
The absence of a symmetric plume could be due to the fact that just one of the two cluster progenitors was a cool core (whose central part still survive at the centre of A1300) similarly massive. The other entropy plume is located to the North and is not compact as the previous one. This could be an indication that A1300 is a complex system in which minor mergers played an important role in the mass assembly and the major merger happened between clusters of different dynamical states.

The comparison with these aforementioned simulations of \citet{2011ApJ...728...54Z}, even if the matching is not perfect, allows us to approximately date the merging of the cluster progenitors: about 3 $(\pm 1)$ Gyr could have passed since when their regions within $\mathrm{R_{200}}$ touched for the first time. Tentatively tracing the previous cores in the entropy map, we can estimate the current projected impact angle which is likely to be $\mathrm{\sim 130\ deg}$ in the plane of the sky. 
This suggests that we are witnessing an almost face-on merging of clusters which allow us to better understand the status of the system related to the projection effects: as in the simulations, the two clusters collide with a relatively low initial impact parameter and after the first core passage the low entropy regions trace the two progenitors. Moreover, the presence of the shock found in the same position of the radio relic suggests that more than 2 Gyr passed since the merging; its weakness ($\mathcal{M}=1.20 \pm 0.10$) could constrain the age of the merging to around 3 Gyr.
Furthermore, the degree of gas mixing as a function of radius (fig.~13 for simulation S2 or S3 in \citealt{2011ApJ...728...54Z}) is very efficient in the off-axis merging, because the cores feel their mutual interaction and are stripped before the final merging.

The disturbed shape of the X-ray entropy, temperature and surface brightness 2-D distributions reflect a high degree of substructure in the cluster, fundamental for the interpretation of different moments of the assembly history.
The surface brightness, in fact, is asymmetric (Fig.~\ref{psf_rec_dens_maps}) and exhibits some excess of emission due to the influence of the filamentary structure through which groups fall. There is a mutual perturbation among the cluster and the infalling groups which keep their identities while they experience the attraction of the cluster, as appears evident from the analysis of the red and blue galaxy fractions as a function of position and luminosity (Appendix A).

The substructures found in A1300 show several episodes of accretion: a filament shows up in the DS-test (Fig.~\ref{dstest}) and in the galaxy distribution (Fig.~\ref{density_maps}) we detect groups entering into the virial region of the cluster (Fig.~\ref{psf_rec_dens_maps}) and over-densities in the inner part of the cluster without an X-ray identification. The last case refers to some peaks that we observe inside $\mathrm{R_{200}}$ by the density map. In fact, when a group enters in a cluster, it is stripped almost immediately of its hot gas \citep{2006PASP..118..517B}. Conversely, group galaxies  can keep orbiting around the cluster centre for some time even after being captured in the deep potential well. Thus, they may retain part of their common DM halo remaining effectively bound together for longer times and appearing as galaxy over-densities. This happens mainly because of the difference in the behaviour of collisional and non-collisional components of clusters or groups (galaxies and ICM respectively) as explained for example by \citet{2002ASSL..272....1S}. 
Moreover, we compared the velocity distributions of the northern and southern components (both at a distance larger than 1 Mpc from the core) of the cluster. The NE filament (in which the Group 4 is embedded, Fig.~\ref{group_snapshots_and_CMD}, corresponding to the northern red circles of the DS test, Fig.~\ref{dstest}) is receding w.r.t. the observer while the southern one (in which is embedded the Group 10, Fig.~\ref{group_snapshots_and_CMD}, corresponding to the southern red circles of the DS test, Fig.~\ref{dstest}) appears approaching to us. In this way we are able to give a three dimensional picture of the mass assembly history using all data available.

The substructure analysis revealed also that the total mass of our X-ray selected groups is around 20\% of that of the cluster. One of them (Group 4, $\mathrm{M_{200}\approx10^{14}\ M_{\sun}}$) is already embedded in the virial region of the cluster (also in terms of gas, as it is possible to see from the X-ray emission in Fig.~\ref{psf_rec_dens_maps}) and contributes alone to half of this value.
This confirms the relation suggested by a recent simulation \citep{2011arXiv1105.1397C}, i.e. that richness of the largest subgroup is typically 20\% of that of its host cluster (even if this relation exhibits a wide scatter). 

Fig.~\ref{BCG_fig} shows the elongated shape of the galaxy distribution in the central region of the cluster, compared with the X-ray emission: the brightest galaxies, together with their satellites, trace the disturbed X-ray emission. 
A bit further from the central emission of the gas we detect groups both in the X-rays and in the optical bands (as galaxy concentration).
The X-ray surface brightness (better investigated in the PSF reconstructed image, Fig.~\ref{psf_rec_dens_maps}) appears elliptical at the centre but a projected filament to the South suggests the direction of a past accretion event. The X-ray surface brightness appears elongated towards the direction of the filaments in which the infalling groups are embedded, thus the ongoing mergers will modify again the shape of the cluster as it happens in the North-East. 

Combining all the available information, A1300 could have been the result of the merging of two clusters of similar mass.  
The collision happened most likely with a small impact parameter (as inferred from the entropy and the temperature maps) and this shock heated the gas, yielding all the features visible in X-rays. 
A1300 still accretes via filaments and groups, which suggests that this cluster is still in the formation process (cf. \citealt{1997MNRAS.291..353B}). Indeed, it could be experiencing a relaxation at the observed time, just after the merging, in which the gas starts settling down and the different components mix. According to our estimations the cluster will increase its total mass about 60\% through filaments (including groups) in the next Gyr with a medium velocity of $\mathrm{\sim2000\ km\ s^{-1}}$.

\section{Conclusions}

We investigated the recent assembly history of the REFLEX-DXL post-merging cluster A1300 through the X-ray and optical (spectroscopy and photometry) data. Our main results are summarized as follows:

\begin{itemize}

\item[$\bullet$] The galaxy distribution of red sequence galaxies reveals a filamentary structure departing from the inner area of the cluster which extends beyond $\mathrm{R_{200}}$. This suggests that galaxies could be pre-processed in the groups that we identified embedded in the filaments surrounding A1300.

\item[$\bullet$] The distribution of the red galaxies of the cluster (mostly concentrated at the centre) exhibits a marked correspondence with the entropy (and temperature) features, suggesting that the galaxies track the information provided by the gas in the central region out to larger cluster-centric radii. The projection effects combined with the status of the merging play an important role in this configuration.

\item[$\bullet$]  The X-ray surface brightness distribution is clearly affected by the presence of filamentary structures through which groups enter the virial radius of the cluster.

\item[$\bullet$] A possible forward shock (consistent with a Mach number $\mathcal{M}=1.20 \pm 0.10$) was identified in the southern part of the cluster coincident with a candidate relic (\citealt{1999MNRAS.302..571R} and \citealt{2011arXiv1102.1901G}) found in the radio bands. Surprisingly also the galaxy distribution presents a sharp edge in the SW.

\item[$\bullet$] A comparison with simulations suggests that 3~($\pm 1$)~Gyr elapsed since the two cluster progenitors (still visible in the entropy map) started merging with a likely projected impact parameter between 462 and 932 kpc.

\end{itemize}

A1300 could have been disturbed by another cluster of similar mass.  
The collision happened most likely with a small impact parameter and this shock heated the gas, yielding all the features visible in X-rays. 
The cluster is now likely to experience a relaxation, just after the merging, in which the gas starts to settle down and the different components mix.

The brightest galaxies of A1300 are coincident with the X-ray peaks in the central region (Fig.~\ref{BCG_fig}) or X-ray signatures of substructure (Fig.~\ref{psf_rec_dens_maps}): these massive galaxies could be reminiscent of former  Brightest Group Galaxy (BGG) that have merged to form the main cluster. It is thus possible to study different stages of a merging even by analysing a single cluster: A1300 is a good example for this purpose as it shows a filament through which it accretes not only individual galaxies but also groups; disrupted ones visible in the optical but not any more in X-rays and other groups still in the process of crossing $\mathrm{R_{200}}$.

\section*{Acknowledgments}
We would like to thank the anonymous referee for the constructive comments, which led to a significant improvement of the quality of this paper. 
FZ acknowledges the support from and participation in the International Max-Planck Research School on Astrophysics at the Ludwig-Maximilians University.
DP acknowledges the kind hospitality at the Max-Planck-Institut f\"{u}r extraterrestrische Physik (MPE).
FZ thanks Olivier Ilbert, Mara Salvato and Francesco Pace for useful discussions.

\appendix

\section{Substructures: X-ray and Optical Analysis}
\label{substructures_optical}
\begin{table*}
\begin{center}
\resizebox{\hsize}{!}{
\begin{tabular}[!t]{cccccccccccccc}
\hline
 ID & RA         & Dec         & $\mathrm{L_X}$ (0.5-2 keV) & $\mathrm{M_{200}^X}$  & $\mathrm{M_{200}^{dyn}}$ & kT              &  $\mathrm{R_{200}}$ & Density & 	$\sigma$	& Red fr. & Blue fr.  & $\mathrm{L_R}$               & mag$_{\mathrm{BCG}}$\\ 
    & deg	 & deg         & $\mathrm{10^{42} erg s^{-1}}$       &$\mathrm{10^{13}M_{\sun}}$   &$\mathrm{10^{13}M_{\sun}}$  & keV   &      arcmin &   $\mathrm{arcmin^{-2}}$ & &   &   & $\mathrm{10^{11}\ L_{\sun}}$ & $\mathrm{mag_{AB}}$ \\
\hline
  4 &  173.03195 & -19.858619  &   $27.71  \pm 2.29 $   &  $13.47 \pm  0.70   $   & $11.70 \pm  2.40$   &  $2.06$&  3.53       & 6.75 &   9.3  &  0.89   &   0.11    &   $\mathrm{2.17 \pm  0.39}$  &   18.11 \\
  5$^\star$ &  172.98374 & -19.921704  & $675.58 \pm 0.85$&$103.99\pm  0.84   $   & $110.00 \pm 20.00$  &  $8.70$&  6.98	& 8.95 &   15.3 &     	0.81   &   0.18    &   $\mathrm{12.13 \pm 1.53} $ &   17.11 \\
  6 &  172.87952 & -19.911543  &   $4.03  \pm 1.09  $   &  $3.92  \pm   0.65  $   & --                  &  $0.94$&  2.34       & 7.10 &   10.3  &     	0.27   &   0.73    &   $\mathrm{1.12 \pm  0.19} $ &   17.87 \\
 10 &  173.10082 & -20.008946  &   $8.43 \pm 1.61   $   &  $6.29  \pm   0.74  $   & $4.37  \pm  2.99$  &  $1.25 $&  2.74	& 5.43 &   5.8  &   	0.56   &   0.44    &   $\mathrm{1.92 \pm  0.22}$  &   17.95 \\
\hline                           

\end{tabular}
   }
\end{center}
\caption{Properties for each group in the field of A1300. Col.1 gives the ID ($\mathrm{ID=5^\star}$ identifies the main body of A1300), col.2 and col.3 the coordinates of the centre, col.4 the X-ray luminosity in the band 0.5-2 keV, col.5 the X-ray mass, col.6 the dynamical mass, col.7 the temperature for each group, col.8 the value of $\mathrm{R_{200}}$ from X-ray analysis (for which we have adopted the same scaling relations of \citealt{2010ApJ...709...97L}), col.9 the number of galaxies per $\mathrm{arcmin^2}$ after the background subtraction, col. 10 the significance w.r.t. the background, col.11 and col.12 the red and blue galaxy fraction, col.13 the total luminosity of the group computed by summing all the single galaxy luminosities in the R band and col.14 the R magnitude of the brightest galaxy found in the group. All the quantities listed in col.4-7 are obtained within a radius $\mathrm{R_{200}}$ while all those in col.9-13 within a radius $\mathrm{R_{500}}$.}
\label{groups_table}
\end{table*}

In the X-ray image of A1300 (Fig.~\ref{psf_rec_dens_maps}), obtained by applying the technique described in \cite{2009ApJ...704..564F}, different regions with extended emission were detected around the main cluster (after removing the point sources) with a significance higher than 4$\sigma$ w.r.t. the background. 
 These regions were compared with the 2-D distribution galaxies in the optical catalogue for the eventual identification of over-densities (groups; marked by green and black ellipses in Fig.~\ref{groups_appr_rec}). 
To assess the significance of these over-densities, we estimate the local density of galaxies at the location of each group and correct for background contamination. We thus select three square regions (highlighted in green in Fig.~\ref{density_maps}) at distances larger than $\mathrm{R_{200}}$ where no significant structures are evident in order to estimate the mean background density.
Four candidate groups were dropped from the sample because the background correction exceeds the estimated density. The remaining groups have a significance of more than 5$\sigma$ above the background. The redshift information is then used to perform an X-ray analysis for all groups. 

In order to assess the membership of the groups to the cluster we perform a second time the background correction using, this time, the cluster photometric member galaxies.
We identify as groups belonging to the cluster all those over-densities with a significance larger than 5$\sigma$ w.r.t. the background. Of these groups we removed number 3 and number 11, since the possibility of contamination from the cluster and/or from their neighbours is high. All remaining groups were classified as associated to the cluster with a low probability (given the uncertainties on our photometric redshifts the analysis is based on a qualitative approach). 
Where available, spectroscopic redshifts were used to confirm their membership to the cluster. 
The background-subtracted surface galaxy density and the significance w.r.t. the background of each group associated to the cluster are shown in Tab. \ref{groups_table} (col.9 and 10, respectively).

\begin{figure*}
\includegraphics[width=0.432\hsize]{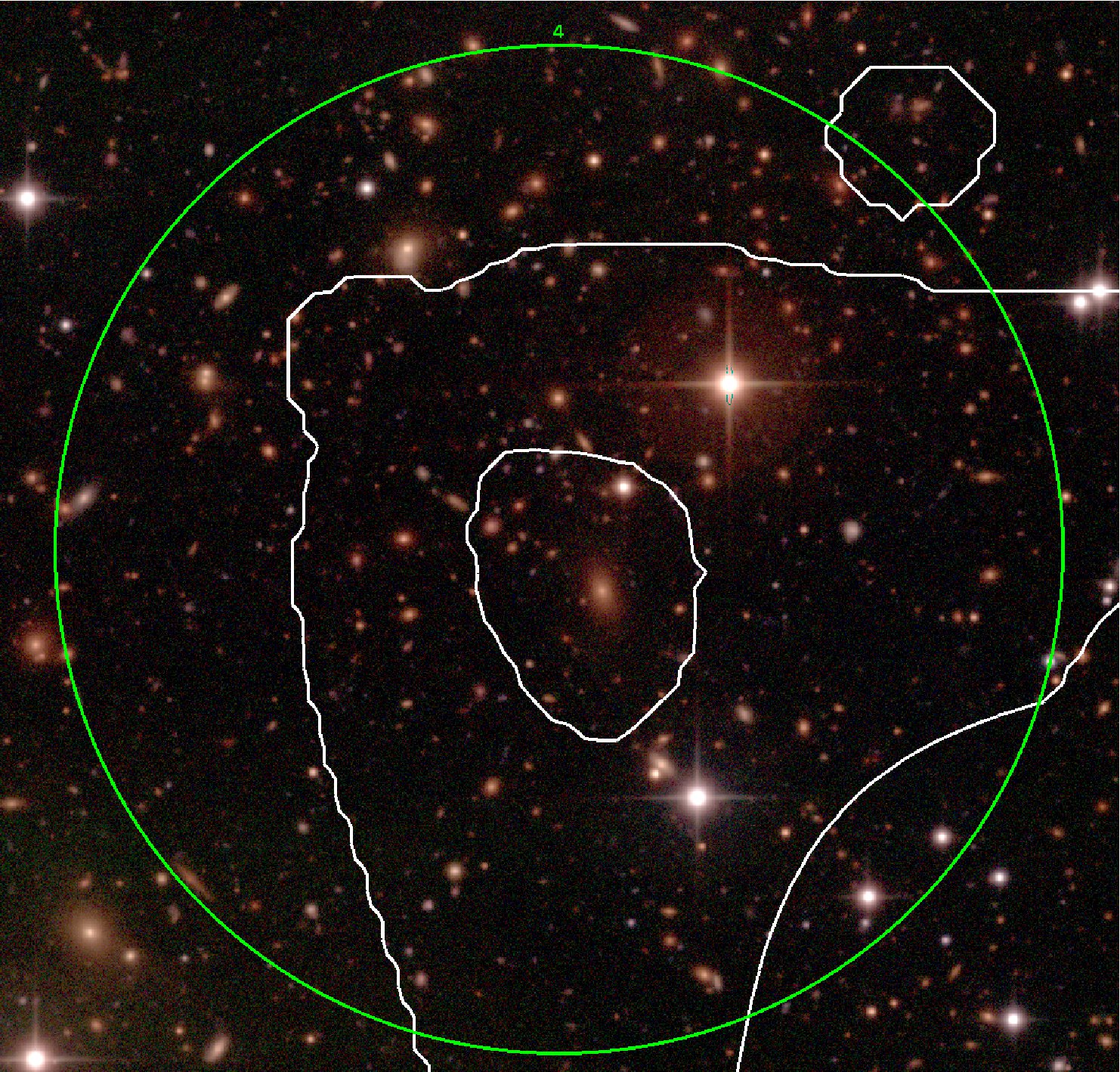}
\includegraphics[width=0.532\hsize]{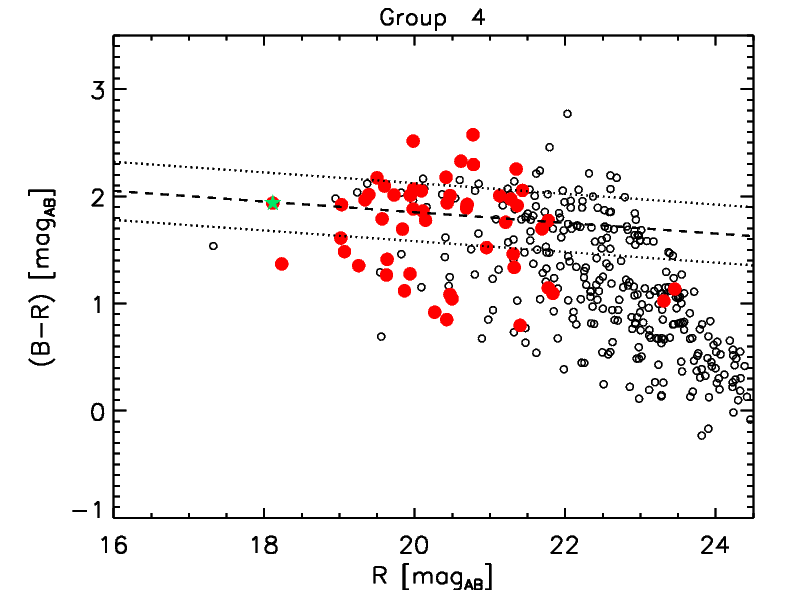}
\includegraphics[width=0.432\hsize]{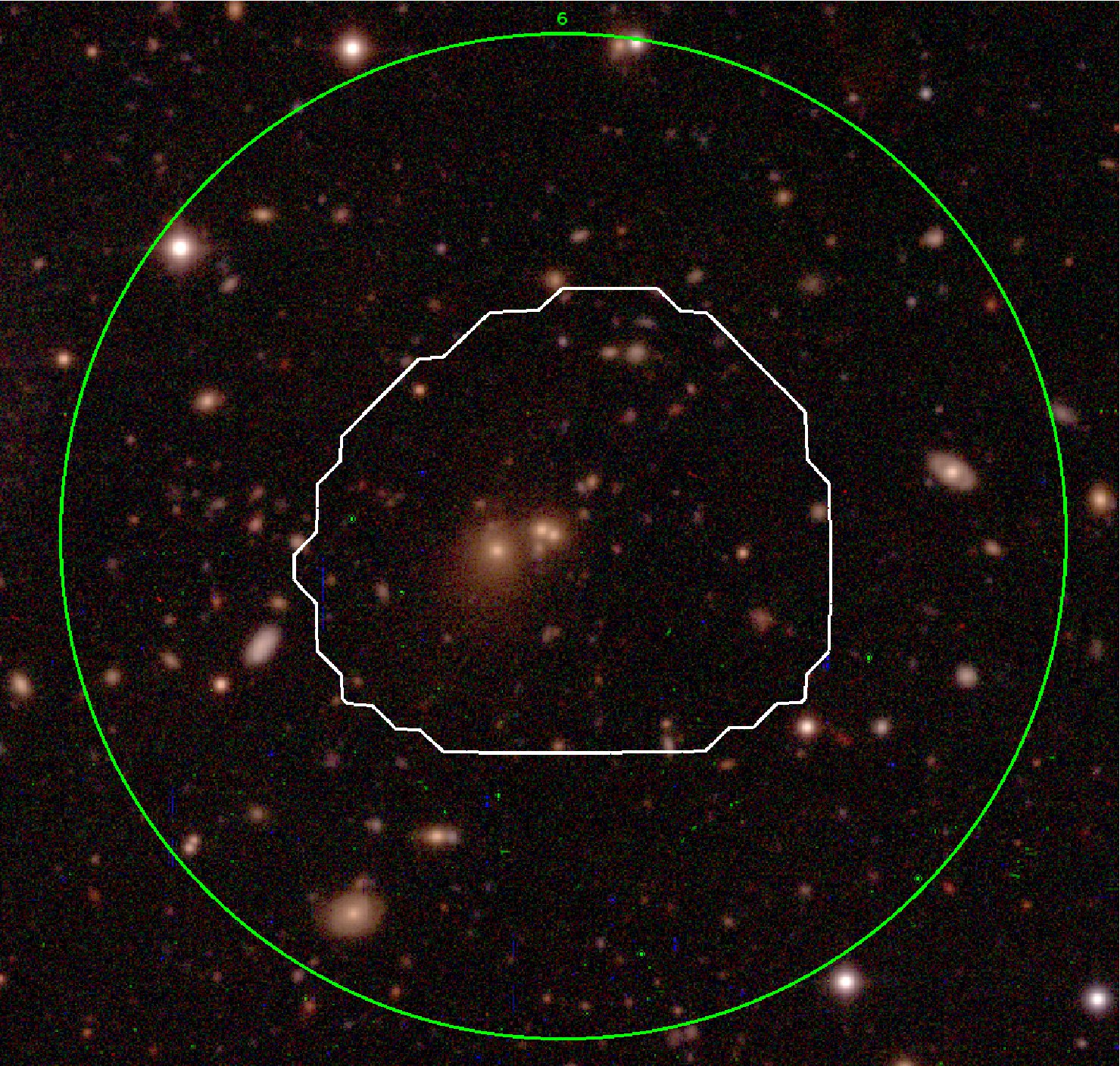}
\includegraphics[width=0.532\hsize]{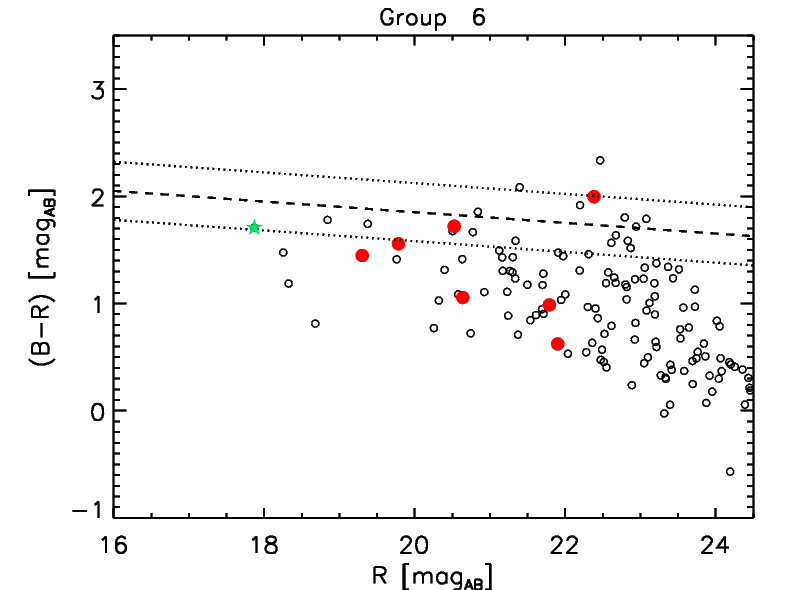}
\includegraphics[width=0.432\hsize]{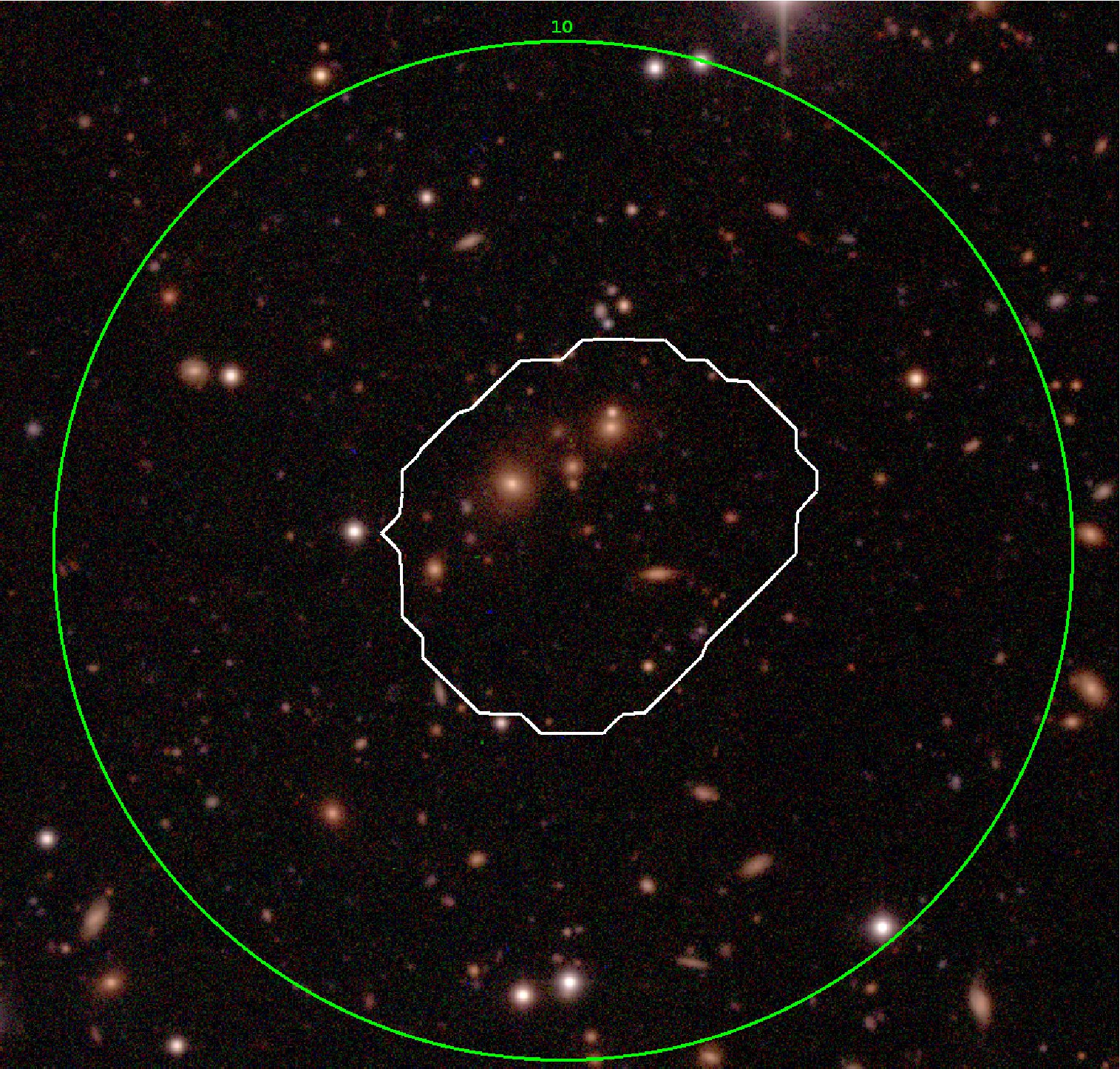} 
\includegraphics[width=0.532\hsize]{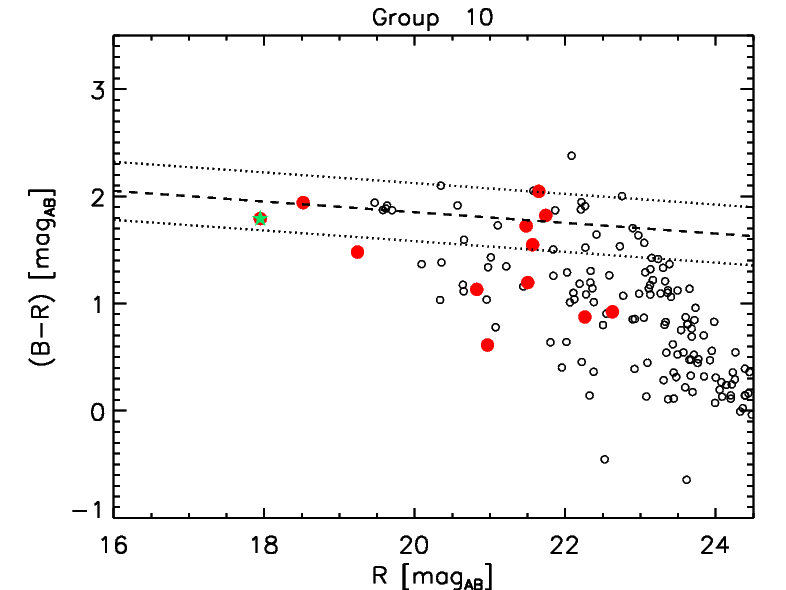}
\caption{On the left: snapshots of group 4, 6 and 10 (from top to bottom respectively); the green circle is the region within $\mathrm{R_{500}}$ and in white are the X-ray emission contours overlaid. A smoothed (with a Gaussian of 8 pixels, or 32") point-source-free signal-to-noise map is used for the contours. The two contour levels have a significance of $\mathrm{3 \times 10^{-16}\ erg\ s\ cm^{-2}\ arcmin^{-2}}$ (which draws the cluster X-ray shape in the top panel) and $\mathrm{1 \times 10^{-14}\ erg\ s\ cm^{-2}\ arcmin^{-2}}$. On the right: colour magnitude diagrams for each group. Black circles represent all photometric members of the cluster comprised within $\mathrm{R_{500}}$ of each group, red dots mark all the galaxies with spectroscopic redshifts, and the green star their BGG. The dashed line marks the best fit for the cluster red sequence the dotted line its $3\sigma$ spread. The catalogue is cut for the magnitude limit of $\mathrm{B=24.9}$ and $\mathrm{R=24.5}$.}
\label{group_snapshots_and_CMD}
\end{figure*}

Cluster membership for each group was estimated using a number of different techniques, as only a fraction of the groups are covered by spectroscopy. For these groups, indeed, we rely upon the information obtained from combining the distribution of cluster photometric members with the presence of a red sequence in the colour-magnitude diagram (Fig.~\ref{group_snapshots_and_CMD}). Visual inspection also confirmed galaxy over-densities and the presence of a central bright galaxy.

All properties derived from this optical analysis are listed in Table \ref{groups_table}, where we computed the density, the fraction of red and blue galaxies and the total luminosity within $\mathrm{R_{500}}$ in order to limit contamination from other groups. Snapshots and colour-magnitude diagrams of the same groups are available in Fig.~\ref{group_snapshots_and_CMD} with X-ray emission contours (in white).

We used the scaling relations of \cite{2010ApJ...709...97L} after deriving the value of $\mathrm{R_{200}}$ for each group and assuming that they lie at the same redshift of the cluster. X-ray properties for groups belonging to the LSS of the cluster are listed in Table \ref{groups_table}. In the computation of the X-ray luminosity $\mathrm{L_X}$, we have taken into account the finite size of the flux extraction area. The full $\mathrm{L_X}$ is estimated based on the observed counts and the expected missed flux, based on the beta-model. This is required in order use the scaling relations which were calibrated for the full $\mathrm{L_X}$.
The masses are estimated based on the measured $\mathrm{L_X}$ and its errors, using 
the scaling relation of \cite{2010ApJ...709...97L} described by their Equation 13:
\begin{equation}
 \frac{\langle M_{200} E(z)\rangle}{M_0} = A  \left( \frac{\langle L_X E(z)^{-1}\rangle}{L_{X,0}} \right) ^\alpha
\end{equation}
where  $E(z)\equiv \sqrt{\Omega_m(1+z)^3+\Omega_\lambda}$ is the Hubble parameter evolution for a flat metric, $\mathrm{ M_0 = 10^{13.7}\ h_{72}^{-1}\ M_{\sun}}$ and $\mathrm{L_{X,0} = 10^{42.7}\ h_{72}^{-2}\ erg \ s^{-1}}$.
The intrinsic scatter in this relation is 20\% (Finoguenov et al. in 
preparation) and it is larger than a formal statistical error associated with 
the measurement of $\mathrm{L_X}$.
We use the L-T relation to compute the temperature, which we used for the 
computation of the k-correction.  Even though we report all the results obtained with the PSF reconstruction method, for the X-ray properties of the main body of A1300 (represented in Tab. \ref{groups_table} by the ID 5) we rely on those derived more accurately in  Z06. The value for the total mass is in agreement with these authors.

Each group hosts at the centre of its X-ray emission a very bright galaxy (identified as BGG), marked as the green star in each CMD in Fig.~\ref{group_snapshots_and_CMD} .
The total luminosity of each group increases with the red galaxy fraction 
and suggests that the bulk of blue galaxies lies mostly at the faint end of the cluster galaxy population \citep{2004ogci.conf..426L}.

In more detail, Group 4 is the most massive and the closest to the cluster among the secondary groups. It shows a red and blue galaxy fraction very similar to that of A1300 and its BGG has a spectroscopic redshift of 0.31. Its galaxy population could have evolved differently in this massive group. The ongoing spectral index analysis will provide a deeper understanding of the actual physical processes in this region (Ziparo et al. in preparation).

Group 6 appears to be crossing $\mathrm{R_{200}}$, however its galaxies are not particularly influenced in terms of their red and blue galaxy fraction. Indeed, it presents a completely different population with respect to the cluster and appears as the densest group (7.1 galaxies per $\mathrm{arcmin^2}$). This over-density is the most significant: its BGG (the most luminous with respect to other groups) coincides with the centre of the X-ray emission and its density clearly stands up at more than 3$\sigma$ above the density of the background. This group could be still not part of the system or it could be so close to the cluster just for a projection effect, due to a possible large scale filamentary structure pointing towards the observer \citep{1997MNRAS.291..353B}. As sufficient spectroscopic information is currently missing, it is difficult to obtain a definitive conclusion about this group. 

Finally, Group 10 is the smallest and least dense. The spectroscopic data confirm that it is embedded in a filament and the optical image (Fig.~\ref{group_snapshots_and_CMD}) shows that the brightest galaxies within the X-ray emission are well aligned in the direction of the cluster (some of them, including the BGG are at the same photometric redshift as the cluster). The red and blue galaxy fractions are similar, suggesting that this group may be still relatively young and may not have experienced strong interactions with the cluster (future studies will address this). The luminosity of this group is modest in the optical but in the X-rays it is ranked as the second most luminous.

\bsp

\label{lastpage}

\end{document}